%% file: Resubmission.tex
\documentclass[journal]{IEEEtran}
\IEEEoverridecommandlockouts
\usepackage{cite}
\usepackage{amsmath,amssymb,amsfonts}
\usepackage{algorithmic}
\usepackage{graphicx}
\usepackage{textcomp}
\usepackage{svg}
\usepackage{graphicx}
\usepackage{xcolor}
\usepackage{booktabs}
\usepackage{multirow}
\usepackage{booktabs}
\usepackage{longtable}
\usepackage{array}
\usepackage{tabularx}
\usepackage{gensymb}
\usepackage{dblfloatfix}
\usepackage{enumitem}

\newlength{\NomLabelWidth}
\input{math_and_tools.tex}

\usepackage{amsthm}
\newtheoremstyle{boldtheorem}
{3pt} 
{3pt} 
{} 
{} 
{\bfseries} 
{.} 
{.5em} 
{\thmname{#1}\thmnumber{ #2}.\thmnote{ \textbf{#3}}} 
\theoremstyle{boldtheorem}

\newtheorem{theorem}{Theorem}

\allowdisplaybreaks

\usepackage{etoolbox}

\def\BibTeX{{\rm B\kern-.05em{\sc i\kern-.025em b}\kern-.08em
    T\kern-.1667em\lower.7ex\hbox{E}\kern-.125emX}}

\begin{document}

\title{Unlocking Transmission Flexibility under Uncertainty: Getting Dynamic Line Ratings into Electricity Markets}

\author{\IEEEauthorblockN{Zhiyi~Zhou,~\IEEEmembership{Graduate Student Member,~IEEE,}
Christoph~Graf, and~Yury~Dvorkin,~\IEEEmembership{Member,~IEEE}}

\thanks{Manuscript received June 30, 2025; revised September 22, 2025; accepted November 17, 2025. This work was supported in part by the
US Department of Energy Advanced Research Projects Agency–Energy under
Grant DEAR000130010.13039/100000001, and in part by National Science
Foundation under Grant OISE 2330450. \textit{(Corresponding author: Zhiyi Zhou)}}
  \thanks{Zhiyi Zhou is with the Department of Civil and Systems Engineering, Johns Hopkins University, Baltimore, MD 21218 USA (email: zzhou124@jh.edu).}
  \thanks{Christoph Graf is with the Institute for Policy Integrity, New York University School of Law, New York, NY 10012 USA, and also with the Program on Energy and Sustainable Development (PESD), Stanford University, Stanford, CA 94305 USA (email: christoph.graf@nyu.edu).}%
  \thanks{Yury Dvorkin is with the Department of Electrical and Computer Engineering and Department of Civil and Systems Engineering, Johns Hopkins University, Baltimore, MD 21218 USA (email: ydvorki1@jhu.edu).}
}  

\maketitle

\begin{abstract}

Static transmission line ratings may lead to underutilization of line capacity due to overly conservative assumptions. Grid-enhancing technologies (GETs) such as dynamic line ratings (DLRs), which adjust line capacity based on real-time conditions, are a techno-economically viable alternative to increase the utilization of existing power lines. Nonetheless, their adoption has been slow, partly due to the absence of operational tools that effectively account for simultaneous impacts on dispatch and pricing. In this paper, we represent transmission capacity with DLRs as a stock-like resource with time-variant interdependency, which is modeled via an approximation of line temperature evolution process, decoupling the impacts of ambient weather conditions and power flow on transmission line temperature and thus capacity. We integrate DLRs into a multi-period DC optimal power flow  problem, with chance constrains addressing correlated uncertainty in DLRs and renewable generation. This yields non-convex problems that we transform into a tractable convex form by linearization. We derive locational marginal energy and ancillary services prices consistent with a competitive equilibrium. Numerical experiments on the 11-zone and 1814-node NYISO systems demonstrate its performance, including impacts on dispatch, pricing, and marginal carbon emissions.
\end{abstract}

\begin{IEEEkeywords}
    Dynamic line ratings, stochastic electricity market, grid enhancing technologies
\end{IEEEkeywords}

\section*{Nomenclature}

\newcommand{\longestitem}{$C_{0,o,g}$}

\vspace{0.5em}
\noindent\textit{Acronyms:}
\begin{ldescription}{\longestitem}
\item [DLR] Dynamic line rating
\item [SLR] Static line rating
\item [GET] Grid-enhancing technology
\item [DC] Direct current
\item [OPF] Optimal power flow
\item [LMP] Locational marginal price
\item [LMRP] Locational marginal reserve price
\item [LME] Locational marginal emission
\item [EMS] Energy management system
\item [CC] Chance constraint
\item [SOC] Second-order cone
\item [CIGRE] Conseil International des Grands Réseaux Électriques
\item [NYPA] New York Power Authority
\item [CAISO] California Independent System Operator
\item [NYISO] New York Independent System Operator
\end{ldescription}

\vspace{0.5em}
\noindent\textit{Sets and Indices:}
\begin{ldescription}{\longestitem}
\item [$\mathcal{V}$] Set of buses, index $i$
\item [$\mathcal{E}$] Set of transmission lines, index $e$
\item [$\mathcal{G}$] Set of generators
\item [$\mathcal{W}$] Set of wind farms
\item [$t$] Time indices

\end{ldescription}

\vspace{0.5em}
\noindent\textit{Parameters:}
\begin{ldescription}{\longestitem}
\item [$q_c$] Convection heat loss (W/m)
\item [$q_r$] Radiated heat loss (W/m)
\item [$q_s$] Solar heat gain (W/m)
\item [$q_J$] Joule heat gain (W/m)
\item [$T_c^\text{max}$] Maximum conductor temperature (\celsius)
\item [$W$] Combined weather condition parameters
\item [$\Delta r, \Delta s$] Ignorable terms during the approximation of line thermal process
\item [$\Delta I, \Delta T$] Difference between maximum values of current/temperature and actual value (A, \celsius)
\item [$\mu^{a},\mu^{b},\mu^{c},\mu^{d}$] DLR parameters calculated by both current ambient conditions and conductor parameters
\item [$w$] Actual wind generation (MW)
\item [$V_c$] Voltage magnitude (MV)
\item [$\rho$] Air density (kg/$\text{m}^3$)
\item [$\rho_r$]  Relative air density compared with that at sea level 
\item [$A$] Rotor swept area ($\text{m}^2$)
\item [$v$] Wind speed (m/s)
\item [$\alpha_s$] Solar absorptivity 
\item [$Q_s$] Total solar radiated heat intensity (W/m$^2$)
\item [$D$] Diameter of conductor (m) 
\item [$T_a$] Ambient temperature $T_a$ (\celsius)
\item [$T_A$] Ambient temperature in absolute scale $T_A = T_a + 273$ (K)
\item [$R_c$] Conductor resistance (DC) at temperature $T_c$ ($\Omega/$m) 
\item [$R_a$] Conductor resistance (DC) at ambient temperature $T_a$ ($\Omega/$m)
\item [$R_\text{max}$] Conductor resistance (DC) at thermal rating $T_{\text{max}}$ ($\Omega/$m)
\item [$\alpha_T$] Temperature coefficient of conductor resistance (\celsius$^{-1}$)
\item [$R_{\text{ref}}$] Conductor resistance (DC) at a reference temperature $T_{\text{ref}}$ ($\Omega/$m)
\item [$T_x$] Difference between conductor and ambient temperatures as $T_x = T_c - T_a$ (\celsius)
\item [$h_r$] Radiated cooling coefficient (W/(m$^2\cdot$K))
\item [$\varepsilon$] Emissivity
\item [$\sigma_B$] Stefan-Boltzmann constant (W/(m$^2\cdot$K$^4$))
\item [$\lambda_f$] Thermal conductivity of air (W/(m$^2\cdot$K))
\item [$N_u$] Nusselt number based on wind speed $v$ 
\item [$N_{\text{Re}}$]  Reynolds number 
\item [$K_{\text{angle}}$]  Wind direction factor as the function of the angle between wind direction and line axis 
\item [$\phi$]  Line axis 
\item [$v_f$]  Kinematic viscosity (m$^2$/s)

\item [$h_c$] Convective cooling coefficient (W/(m$^2\cdot$K))

\item [$\Sigma_{\omega \xi}$] Joint covariance matrix for $\boldsymbol{\omega}$ and $\boldsymbol{\xi}$
\item [$\Sigma_{\omega \varsigma}$] Joint covariance matrix for $\boldsymbol{\omega}$ and $\boldsymbol{\varsigma}$

\item [$\gamma$] Gradient of renewable generation/DLR parameter with respect to ambient conditions, showing sensitivity of the ambient condition forecast errors
\item [$\Gamma_{\omega\xi}$] Matrix consist of $\gamma_{w}$ and $\gamma_{f}$ with $(\Gamma_{\omega\xi})_i = \gamma_{w,i},(\Gamma_{\omega\xi})_e = \gamma_{f,e}$
\item [$\Gamma_{\omega\varsigma}$] Matrix consist of $\gamma_{w}$ with $(\Gamma_{\omega\varsigma})_i = \gamma_{w,i}$
\item [$\epsilon$] Maximum probability of constraint violations
\item [$\sigma_{le}$] Standard deviation of stochastic DLR, line $e$
\item [$b_{\omega e, k}$] Covariance between $\boldsymbol{\omega}_k$ and $\boldsymbol{\xi}_e$ or $\boldsymbol{\varsigma}_e$
\item [$d$] Load (MW) 
\item [$p_i^{\text{max}}$] Rated power of generator $i$ (MW)
\item [$S_{e,i}$] Power transmission distribution factor (PTDF), equal to flow through line $e$ caused by a unit injection at bus $i$
\item [$c_{i,1}, c_{i,2}$] First-order/Second order cost coefficients of generator $i$ (\$/MWh, \$/MWh$^2$)
\item [$U_i^{\text{up}}, U_i^{\text{dn}}$] Ramp-up/Ramp-down rate for generator $i$ (MW/h)
\item [$\kappa$] Parameters derived from first-order Taylor expansion for approximation of transient temperature revolution process

\end{ldescription}

\vspace{0.5em}
\noindent\textit{Variables:}
\begin{ldescription}{\longestitem}
\item [$T_{e,t}$] Conductor temperature for line $e$ at time $t$ (\celsius)
\item [$f_{e,t}$] Power flow on Line $e$ at time $t$ (MW)
\item [$\boldsymbol{\omega}$] Stochastic wind generation forecast error (MW)
\item [$\boldsymbol{\xi}$] Stochastic line rating forecast error (MW)
\item [$\boldsymbol{\varsigma}$] Multiple stochastic ambient conditions forecast error
\item [$\boldsymbol{\Omega}$] System-side wind generation forecast error (MW)
\item [$\hat{w}$] Wind generation forecast (MW)
\item [$\hat{\mu}^a,\hat{\mu}^b,\hat{\mu}^c,\hat{\mu}^d$] DLR parameters forecast
\item [$p_{i,t}$] Power output of generator $i$ (MW)
\item [$\alpha_{i,t}$] Balancing participation factor of generator $i$
\item [$R_{i,t}^\text{up}, R_{i,t}^\text{dn}$] Up/down reserves for generator $i$
\item [$R_{e,t}^\text{th}$] Auxiliary thermal reserve for line $e$

\item [$\lambda$] Dual variable 

\item [$\pi_{i,t}$] Locational marginal price for bus $i$ (\$/MWh)
\item [$\tau_{i,t}$] Locational marginal reserve price for bus $i$ (\$/MWh)

\end{ldescription}

\IEEEpeerreviewmaketitle

\section{Introduction}\label{A}

The backbone transmission system was not originally designed for distributed and intermittent generation, and while this generation may alleviate congestion in some areas, it may also impose new constraints. Transmission congestion costs in the U.S.\ raised to \$11.5 billion in 2023 \cite{Report:CongestionCost2}. In 2024, CAISO alone curtailed 3.2 million MWh of renewable generation due to limited transmission capacity \cite{Report:CAISOcurtail}.  More recently, in addition to challenges from renewable integration, large loads (e.g., data centers and hydrogen electrolyzers) have also begun to contribute to congestion. The commonly used static line ratings (SLRs), which are derived using conservative \cite{DOE2019DLR} or average \cite{SLR_cal2} assumptions about the operation environment, are shown to underutilize transmission infrastructure up to 30\% even under favorable operational conditions \cite{EPRIreport}. This highlights the need to safely leverage transmission flexibility to enhance utilization of existing transmission infrastructures.

Dynamic line ratings (DLRs) have emerged as a promising grid-enhancing technology (GET) to increase existing transmission capacity without new line construction, which is time-consuming and capital intensive. DLRs have been recognized as a relatively low-cost and fast-to-deploy option, requiring mainly software integration and limited sensor installations \cite{DOE2019DLR,SLR_cal2}. Recent regulatory developments, such as FERC Order No. 881 \cite{FERCOrder881}, further highlight the importance of incorporating ambient-adjusted and dynamic ratings into operational practice.
DLRs adjust line ratings based on real-time weather conditions, and have moved beyond the pilot stage in many regions, becoming part of standard grid operations. The Oncor Electric Delivery Company achieved up to 12\% annual average increases in transmission capacity, which peaked at  30\% under  favorable conditions \cite{DLR:DOEreport}. Similarly, DLR implementation on critical corridors in NYPA yielded capacity increases of up to 15\% during winter \cite{DLR:NYPA}. In Europe, the FLEXITRANSTORE project across six countries increased cross-border transmission capacity by up to 20\% \cite{DLR:EU}. 

While international adoption has advanced---particularly in systems with simplified market mechanisms using system-wide/zonal pricing followed by congestion management and redispatch to ensure secure real-time operations \cite{Graf25}---nodal markets in the U.S.\ may present barriers to broader DLR integration, along with several other limiting factors in management and operational complexities:
\begin{itemize}
    \item [(a)] Implementation complexity. Widespread deployment of DLR requires reliable real-time data on ambient conditions (e.g., wind speed, direction, solar radiation, ambient temperature) and conductor temperatures. This entails installing field sensors, remote monitoring units, and communication infrastructure \cite{DLR:hard1_hardware:}, plus integration into energy management systems (EMS) and market clearing platforms, which introduces additional ``soft'' integration costs. Additionally, interoperability of heterogeneous sensor technologies and  cybersecurity requirements introduce implementation hurdles \cite{attack}. 
    \item[(b)] Reliability concerns. DLRs introduce an additional source of uncertainty in line capacity when applied in look-ahead scheduling while under stressed conditions. Forecasting errors in ambient conditions or unexpected weather changes may lead to capacity overestimation, with potential implications on reliability if coincides with peak demand or adversarially correlated with weather-dependent renewable generation \cite{DLR:hard2_reliability}. 
    \item[(c)] Regulatory and operational integration. Even when DLRs are technically feasible, their integration into market operations requires modified procedures for dispatch, scheduling, and reliability coordination. For example, FERC Order No. 881 mandates the use of ambient-adjusted ratings and calls for the development of frameworks to incorporate DLRs into operational practice \cite{FERCOrder881}. However, system operators must still address how to update ratings in real time, how frequently to refresh forecasts, and how to ensure procedural transparency and accountability for all market participants. 
    \item[(d)] Economic incentives. In nodal markets such as those in the U.S., transmission owners are often regulated entities whose revenues are decoupled from short-term market efficiency gains. Since the benefits of DLRs (lower congestion costs, improved utilization of generation assets) accrue largely to market participants and consumers rather than transmission owners, the incentives for utilities to invest in DLRs are often weak or misaligned \cite{DLR:hard3_market}. 
\end{itemize}
While SLRs inaccurately approximate the actual transmission capacity, which in turn hinders economic, reliability, and environmental gains \cite{Report:CongestionCost2}, moving to DLRs also requires changes to electricity pricing, which guides investment and operational decisions. Locational marginal pricing (LMP) remains one of the predominant approaches in organized electricity markets \cite{pricing3}, composed with the price for energy, transmission congestion, and transmission losses. These prices are derived from the market-clearing model, which  usually use a linear DC approximation. Notably, DLRs will affect the congestion component directly and the loss and energy components indirectly. By adding time-varying capacity, DLRs introduce new sources of variability into nodal prices, which calls for more advanced methods of uncertainty management. Existing approaches to manage uncertainty focus on the demand and renewable generation and include scenarios \cite{scenario1}, robust optimization \cite{robust1}, and chance constraints (CC) \cite{CC:paper1}. CC methods align with industry practices, providing transparent risk management without excessive conservatism of deterministic approaches with fixed security margins. Studies \cite{DLR:PierrePinson} and \cite{DLR:wang} considered the uncertainty in DLRs computations, but ignored important correlations between DLRs and renewable generation. Since weather conditions impact both DLRs and renewable generation, ignoring this correlation may lead to suboptimal dispatch. Prior work \cite{Group:yury} developed electricity pricing with  CC, but without considering DLRs.

This paper models transmission flexibility, enabled by DLRs, and integrates it with electricity pricing. Unlike previous models, i.e., with SLRs \cite{CC:paper3}, DLRs \cite{DLR:bolun}, with steady-state temperature assumptions underlying transmission line limits \cite{DLR:approxi}, we propose a transient
DLR model, a multi-period line rating formulation that explicitly accounts for conductor thermal inertia, capturing how past power flows influence future line capacity. This extension enables the operator’s decisions to directly affect subsequent line ratings, providing additional scheduling flexibility.

The main contributions of this paper are as follows:
\begin{itemize}
\item To incorporate transient DLRs into  routines used for power system and market operations, this paper approximates  transmission line temperature evolution  and itemizes the contributions of ambient weather conditions and power flow to DLR computations under uncertainty. Through linearization, we represent the uncertain state variables, i.e., line temperature and power flow, as affine functions of optimal power flow (OPF) decisions, which relate the uncertain system states to control inputs. 

\item Based on this approximation, this paper incorporates transient DLRs into the multi-period CC DC-OPF framework, which accounts for the correlation between  transmission  and weather-dependent renewable generation.

\item This paper adopts marginal cost-based electricity pricing and considers reserve deliverability under transient DLRs. We also prove electricity market equilibrium with DLRs and analyze the versatile impacts of DLRs on marginal electricity and reserve prices and marginal emission rates. 

\end{itemize}

\section{Conceptual Framework}

In conventional practice, DLRs are computed mainly based on steady-state thermal limits according to ambient weather forecasts under the assumption that the conductor is operated in a thermal steady state. The resulting rating is a deterministic function of the forecasted weather conditions and remains independent of system scheduling and dispatch decisions. In this paper, we extend this concept by explicitly modeling the transient thermal process of overhead conductors as a function of both ambient weather conditions and power flows. By doing so, the proposed transient DLR formulation transforms the inherently inflexible steady-state ratings into flexible ratings that evolve with system operation, thereby providing additional operational flexibility beyond conventional DLRs. Note that the proposed transient DLR does not always offer higher line ratings relative to SLRs and steady-state DLRs. In adverse conditions such as hot and windless weather, the transient model may in fact yield lower line ratings in order to prevent conductor overheating. The key contribution of the proposed transient DLR framework is therefore not in \textit{de facto} guaranteeing increased transfer capability, but rather in verifiably providing a more accurate and flexible representation of thermal line limits. By explicitly accounting for the conductor thermal inertia and the prevailing ambient conditions, the model ensures that line ratings reflect realistic operating limits, whether they are higher or lower than the conventional values.

\begin{figure}[t]
    \centering
    \includegraphics[width=\columnwidth]{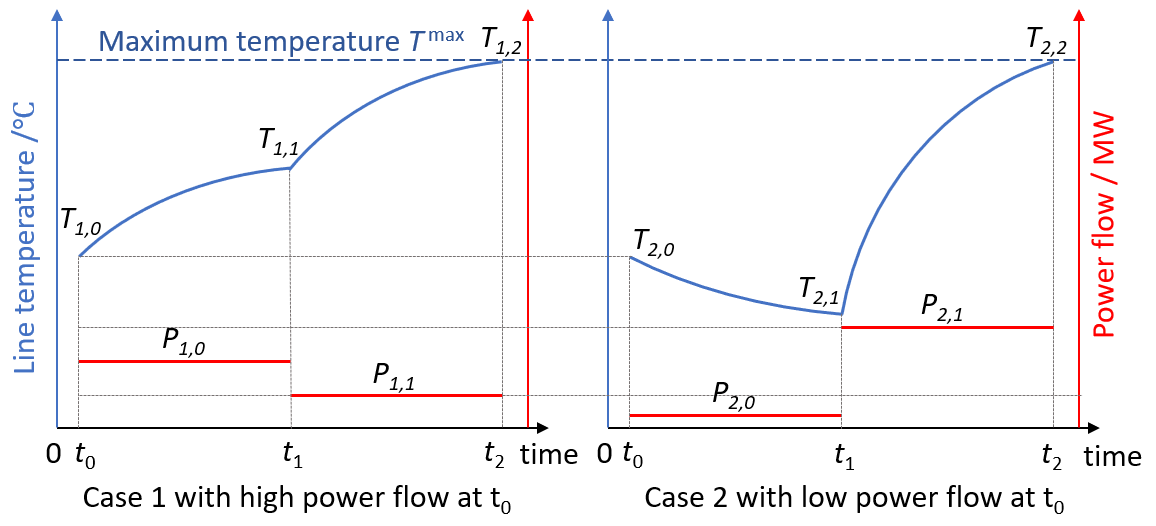}
    \caption{Illustrative cases for transient DLRs: both the power flow trajectories $\{P_{1,0},P_{1,1}\}$ and $\{P_{2,0},P_{2,1}\}$ are valid realizations of the transient DLR over the two time periods, since both cases reach the thermal limit at time $t_2$}
    \label{fig:dynamicTem}
\end{figure}

For a longer operating horizon (e.g., one day), when multi-period thermal dynamics need to be taken into account, the line rating should be defined as a set of power flow sequences that lead the conductor temperature to reach its thermal limit at some critical time $t^*$. In this setting, if the power flow  increases at any instant prior to $t^*$--even if the conductor temperature at that moment is still below the limit--the cumulative heating effect would eventually cause the temperature to exceed the thermal threshold at $t^*$. As illustrated in Fig. \ref{fig:dynamicTem}, Case 1 and Case 2 are subject to identical ambient conditions and have the same initial temperature at $t_0$, i.e., $T_{1,0} = T_{2,0}$. Under different power flow trajectories, $\{P_{1,0},P_{1,1}\}$ and $\{P_{2,0},P_{2,1}\}$, both cases reach the thermal limit at time $t_2$. Hence, both sequences are valid realizations of the transient DLR over the considered horizon. Obviously, the number of such sequences is potentially infinite. This motivates the need to develop an accurate yet compact modeling approach to represent such transmission flexibility and to select the sequence that best serves a given objective function.

\begin{figure}[t]
    \centering
    \includegraphics[width=\columnwidth]{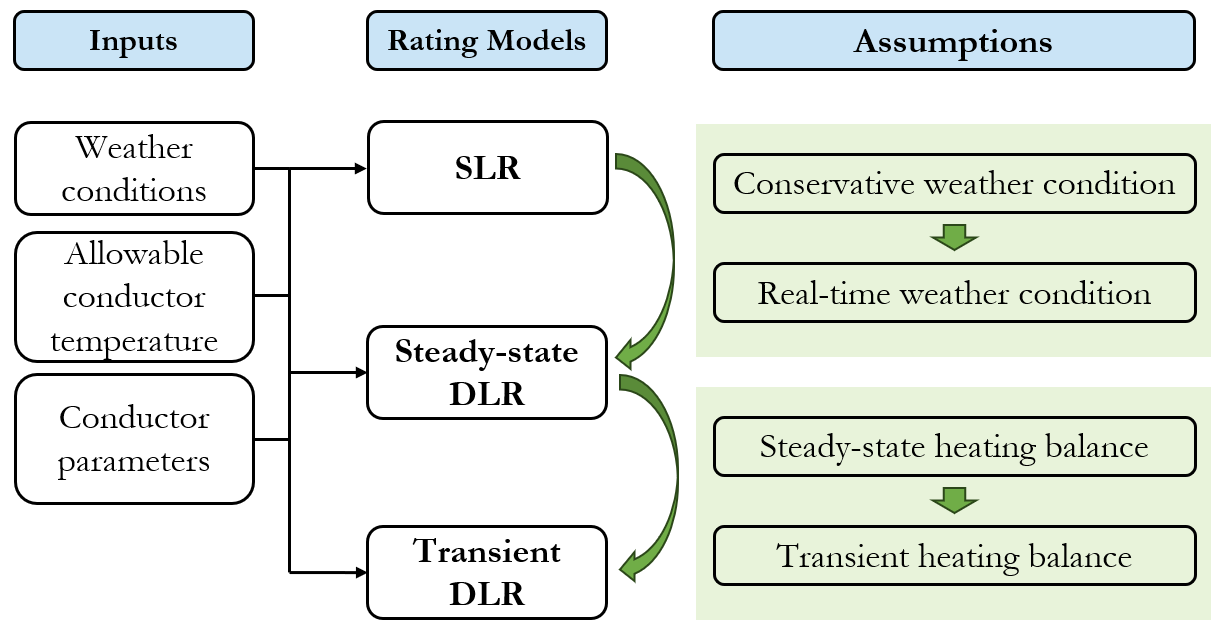}
    \caption{Summary of three line rating models with the main assumptions.}
    \label{fig:three_LR}
\end{figure}

Fig. \ref{fig:three_LR} provides a high-level comparison between SLRs, steady-state DLRs, and the proposed transient DLRs. All models share the same basic inputs: ambient conditions, allowable conductor temperature, and conductor parameters. The key distinctions lies in that SLR relies on conservative weather assumptions to yield a fixed line rating. Steady-state DLR replaces this with real-time weather observations but still models steady-state thermal balance. Transient DLR further advances this modeling by adopting a more accurate transient heating balance, which accounts for conductor thermal inertia.

\vspace{0.6cm}

\section{Uncertainty-Aware Transmission Flexibility}\label{B}
\vspace{0.2cm}

\subsection{DLRs and transmission flexibility}

DLRs are computed based on current ambient conditions (e.g., wind speed, wind direction, ambient temperature, and solar radiation) and conductor properties (e.g., diameter, material, and resistor). This relationship is driven by the steady-state heating balance equation:
\begin{equation}\label{bal}
    q_c(T_c)+q_r(T_c)=q_s+q_J(T_c, f_c),
\end{equation}
where $q_c$ and $q_r$ denote convection and radiated heat losses, both dependent on conductor temperature $T_c$, $q_s$ is solar heat gain, $q_J$ is Joule heat gain determined by temperature $T_c$ and line flow $f_c$. Using (\ref{bal}), we express $f_c$ as a function of $T_c$ and weather conditions $W$ as $f_c = g(T_c | W)$. Then, DLRs are set based on maximum temperature $T_c^\text{max}$:
\begin{equation}\label{DLR_max}
    f_c^\text{max} = g(T_c^\text{max}| W),
\end{equation}
where $g(\cdot)$ denotes the mapping from $T_c^\text{max}$ to $f_c^\text{max}$ under given ambient conditions $W$ at a steady-state equilibrium. The detailed formulation for $g(\cdot)$ is available in Appendix \ref{app:detailed DLR}.

Thermal transient process of a transmission line typically lasts on the order of tens of minutes, with studies reporting thermal inertia constants in the range of 15–60 minutes depending on conductor geometry \cite{INL_ThermalInertia, DLR:TransientAnalysis}. In contrast, the operational time steps in power system scheduling are much shorter: real-time dispatch is executed every 5 minutes across all U.S. ISO/RTOs \cite{FERCOrder825_RT}, while some ISOs are also gradually adopting 15-minute intervals in day-ahead scheduling \cite{CAISO15min_DA}. This thermal inertia enables the power flow to temporarily exceed its deterministic rating $f_c^\text{max}$ derived in \eqref{DLR_max}, without causing the conductor temperature to surpass its upper limit $T_c^\text{max}$. To capture this effect, we further consider the transient heating balance, assuming that weather conditions remain known in each time period:
\begin{equation}
    q_c(T_c)+q_r(T_c)+C\frac{dT_c}{dt}=q_s+q_J(T_c, f_c),\label{trans}
\end{equation}
where $C$ denotes the total heat capacity of the conductor. 

However, since in expression (\ref{trans}), $T_c$ is defined implicitly and cannot be expressed in a closed-form expression, which obstructs modeling an explicit upper bound $T_c \leq T_c^\text{max}$ as a constraint in the OPF problem. Instead, we derive the following approximation in Theorem \ref{theorem:DLR-tran}:

\begin{theorem}[Conservative temperature evolution based on transient heating balance]\label{theorem:DLR-tran}
    Consider the model in (\ref{trans}). Let the current ambient conditions (e.g., wind speed, wind direction, ambient temperature, and solar radiation) and conductor parameters (e.g., diameter, resistor) satisfy:
    \begin{subequations}\label{suffi}
    \begin{equation}
        \Delta r \frac{\partial F_1(\Delta r=0, \Delta s=0)}{\partial \Delta r} \!+\! \Delta s \frac{\partial F_1(\Delta r=0, \Delta s=0)}{\partial \Delta s}\!<\! 0 \label{suffi:F1}
    \end{equation}
    \begin{equation}
        \Delta T \frac{\partial F_2(\Delta T\!=\!0, \Delta I\!=\!0)}{\partial \Delta T} \!+\! \Delta I \frac{\partial F_2(\Delta T\!=\!0, \Delta I\!=\!0)}{\partial \Delta I} \!< \!0 \label{suffi:F2},
    \end{equation}
    \end{subequations}
where $\Delta r$ is a small term associated with $q_r$, which depends on $T_c$ and ambient conditions $W$; $\Delta s$ is a small term related to $q_s$; $F_1(\Delta r, \Delta s)$ quantifies the residual in (\ref{bal}) resulting from the omission of $\Delta r$ and $\Delta s$. $\Delta T = T_c^\text{max}-T_c$ is  the gap between the maximum temperature and actual temperature, $\Delta I = \frac{f_c^\text{max}-f_c}{V_c}$ is the difference between the maximum current and actual current and $V_c$ is the voltage magnitude; $F_2(\Delta T, \Delta I)$ characterizes the effect of partially neglecting $\Delta T$ and $\Delta I$ in $q_r$ and $q_J$ on (\ref{trans}). Then power flow $f_t$ and temperature $T_{t}$ at time $t$, and temperature $T_{t+1}$ at time $t+1$ satisfy:
\begin{equation}\label{DLR:transient}
    T_{t+1} < \mu^a_t + \mu^b_t T_t + \mu^c_t f_t^2 + \mu^d_t f_t^4,
\end{equation}
where $\mu^a_t$, $\mu^b_t$, $\mu^c_t$ and $\mu^d_t$ are parameters calculated by both current ambient conditions and conductor parameters.
\hfill \qed
\end{theorem}

Proof of Theorem \ref{theorem:DLR-tran} is given in Appendix 
\ref{app:DLR-tran}. 

\begin{figure}[t]
    \centering
    \includegraphics[width=\columnwidth]{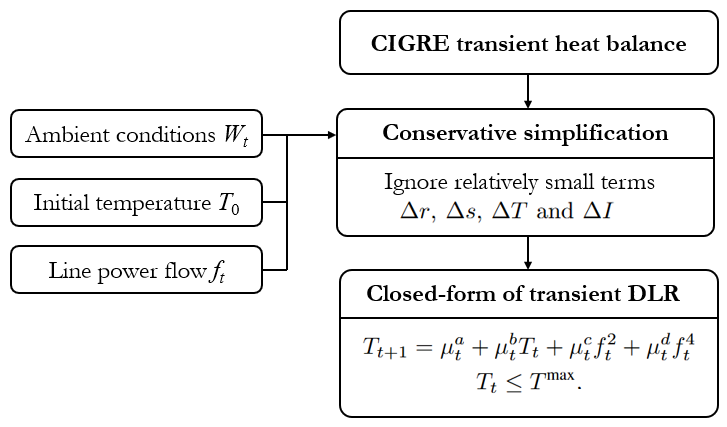}
    \caption{Flowchart for the proposed transient DLR model}
    \label{fig:DLR_flowchart}
\end{figure}

Relative to \eqref{DLR:transient},  computing $T_{t+1}' = \mu^a_t + \mu^b_t T_t + \mu^c_t f_t^2 + \mu^d_t f_t^4$ leads to $T_{t+1}<T_{t+1}'$. When we set $T_{t+1}' \leq T^\text{max}$, it also holds for the actual $T_{t+1}$ as $T_{t+1} \leq T^\text{max}$. Then we can get a conservative mapping function from $f_t$ and $T_t$ to $T_{t+1}$ given weather conditions, decomposing the influences of $f_t$ and $T_t$. The initial static power flow constraint $f_t \leq f^{\text{max}}$ is transformed into a transient line temperature constraint:
\begin{subequations}\label{new_cons}
    \begin{gather}
        T_{t+1} = \mu^a_t + \mu^b_t T_t + \mu^c_t f_t^2 + \mu^d_t f_t^4 \label{new_cons1}\\
        T_t \leq T^\text{max},
    \end{gather}
\end{subequations}
which ensures thermal security and relates to OPF decisions via variable $f_t$. By capturing thermal evolution, \eqref{new_cons} endows a transmission line with stock-like characteristics, which exhibits intertemporal flexibility. Thus, higher power flows increase $T_t$, reducing future flow capacity and incurring an opportunity cost. Similarly, when  power flows are low enough, i.e.,  heat gain is less than convection and radiated heat loss, $T_t$ will  decrease, with increased future transmission capacity. The overall procedure of deriving the transient DLR model is summarized in Fig. \ref{fig:DLR_flowchart}.

Model \eqref{new_cons1} requires both an initial condition and boundary inputs. For real-time conductor temperature tracking applications, initial temperature $T_0$ is set to a given measured conductor temperature, while for OPF problems it is initialized at the steady state value corresponding to the forecasted ambient conditions and power flow at $t=0$. The boundary conditions consist of the exogenous ambient parameter sequence, which determines the coefficients $\{\mu_t^a,\mu_t^b,\mu_t^c,\mu_t^d\}$, and the sequence of line flows $\{f_t\}$, either observed or optimized, subject to the path constraint $T_t \leq T^{
\text{max}},\forall t$.

For each transmission line we use a single set of aggregated weather parameters. Although weather conditions can vary significantly along a transmission corridor \cite{GentleDLRForecasting}, the accuracy limitations can be alleviated by introducing uncertainty model, which we describe in the next section.

Compared with SLR and steady-state DLR models as in \cite{DLR:PierrePinson, DLR:approxi}, the proposed transient DLRs capture conductor thermal inertia, effectively modeling a stock-like property that links ratings across time periods. This enables more flexible and accurate inter-temporal utilization of transmission assets, though it also requires additional hardware and more advanced operational practices.

Compared with more complex numerical models of the full dynamic thermal process as in \cite{DLR:TransientAnalysis, DLR:transient2}, our simplified formulation offers a closed-form representation that decouples the influence of line flows and ambient conditions, making it able to directly embed into OPF framework. The trade-off, however, is that our approximation inevitably sacrifices some accuracy. To mitigate its influence, the simplifications are designed to be conservative, as established in Theorem 1, thereby ensuring that thermal safety is not compromised.

\subsection{Correlated weather-dependent renewable generation and DLR uncertainty}
Weather conditions impact both DLRs and weather-dependent renewable  generation. Neglecting this correlation may lead to erroneous dispatch outcomes \cite{Abboud2021ConcurrentCooling}. In this section we focus on wind power. The same methodology can be readily applied to other weather-dependent resources such as solar PV, hydropower, or concentrating solar power. We first consider correlated wind power and steady-state DLR uncertainty. The uncertain wind power generation at node $i$ and steady-state DLR for line $e$ at time $t$ are as follows:
\begin{subequations} 
    \begin{gather}
        \hat{w}_{i,t} = w_{i,t} + \boldsymbol{\omega}_{i,t} \label{omega_wind}\\
        \hat{f}_{e,t}^\text{max} = f_{e,t}^\text{max} + \boldsymbol{\xi}_{e,t},
    \end{gather}
\end{subequations}
where $\hat{w}_{i,t}$ and $\hat{f}_{e,t}^\text{max}$ denote the stochastic wind generation and steady-state DLR, each composed of deterministic forecast $(w_{i,t},f_{e,t}^\text{max})$ and addictive error $(\boldsymbol{\omega}_{i,t},\boldsymbol{\xi}_{e,t})$. 

We consider ambient parameters that affect both DLR and wind power, i.e., wind speed, direction, and temperature, and denote their forecast errors by random variables $\boldsymbol{\varsigma}_{i,t}$. Both wind generation and steady-state DLRs are nonlinear functions of these ambient parameters. Since the forecast errors of ambient parameters are typically small, the nonlinear functions can be approximated by their first-order Taylor expansions around the forecast values. Consequently, the stochastic derivations $\boldsymbol{\omega}_{i,t}$ and $\boldsymbol{\xi}_{e,t}$ are represented as linear combinations of $\boldsymbol{\varsigma}_{i,t}$: 
\begin{equation}
    \boldsymbol{\omega}_{i,t} = \gamma_{w,i,t}^\top \boldsymbol{\varsigma}_{i,t},\quad\boldsymbol{\xi}_{e,t} = \gamma_{f,e,t}^\top \boldsymbol{\varsigma}_{e,t},
\end{equation}
where $\gamma_{w,i,t}$ and $\gamma_{f,e,t}$ are obtained as gradients:
\begin{subequations}
    \begin{gather}
        \gamma_{w,i,t} = \nabla_{W_{i,t}} w(W_{i,t})\\
        \gamma_{f,e,t} = \nabla_{W_{e,t}} g(T_c^\text{max},W_{e,t}),
    \end{gather}
\end{subequations}
where $W_{e,t}$ denotes the vector of ambient conditions, $w(W_{i,t})=\frac{1}{2}\rho_{i,t} A v_{i,t}^3$ represents wind power generation with air density $\rho_{i,t}$, rotor swept area $A$, and wind speed $v_{i,t}$ \cite{wind_gen}. Similar linearization techniques for uncertainty modeling have been adopted in previous studies \cite{DLR:PierrePinson}, \cite{DLR:forecast}. 

Given the covariance matrix $\Sigma_{\varsigma}$ among all random variables $\boldsymbol{\varsigma}$, the joint covariance matrix $\Sigma_{\omega \xi}$ for $\boldsymbol{\omega}_{i,t}$ and $\boldsymbol{\xi}_{e,t}$ is:
\begin{equation}
    \Sigma_{\omega \xi,t} = \Gamma_{\omega \xi,t}^\top \Sigma_{\varsigma} \Gamma_{\omega \xi,t},
\end{equation}
where $\Gamma_{\omega \xi,t}$ assembles the corresponding sensitivity vectors in alignment with $\boldsymbol{\varsigma}_{i,t}$ and $\boldsymbol{\varsigma}_{e,t}$:
\begin{equation}
    (\Gamma_{\omega \xi,t})_i = \gamma_{w,i,t},\quad (\Gamma_{\omega \xi,t})_e = \gamma_{f,e,t}.
\end{equation}

Similarly, the uncertainty in temperature evolution for transient DLR is:
\begin{equation}
    \hat{\mu}^{[\cdot]}_{e,t}(\boldsymbol{\varsigma}_{e,t}) = \mu^{[\cdot]}_{e,t} + \gamma_{e,t}^{{[\cdot]}\top} \boldsymbol{\varsigma}_{e,t}, \label{omega_DLR}
\end{equation}
where $\mu^{[\cdot]}_{e,t}$ denotes the transient DLR parameters in \eqref{DLR:transient}, and $[\cdot]\in\{a,b,c,d\}$. The covariance matrix $\Sigma_{\omega \varsigma,t}$ for $\boldsymbol{\omega}_{i,t}$ and $\boldsymbol{\varsigma}_{e,t}$ is:
\begin{equation}
    \Sigma_{\omega \varsigma,t} = \Gamma_{\omega \varsigma,t}^\top \Sigma_{\varsigma} \Gamma_{\omega \varsigma,t},
\end{equation}
where $\Gamma_{\omega \varsigma,t}$ arranges the sensitivity vectors $\gamma_{w,i, t}$ in the appropriate columns aligned with $\boldsymbol{\varsigma}_{i,t}$.

Building upon transient DLR constraints in (\ref{new_cons}) and uncertainty model in (\ref{omega_wind}) and (\ref{omega_DLR}), we construct a stochastic DLR formulation embedded in a CC framework:
\begin{subequations}\label{Temp_CC}
\begin{align}
    \hat{T}_{e,t+1} =& \hat{\mu}^a_{e,t}(\boldsymbol{\varsigma}_t) + \hat{\mu}^b_t(\boldsymbol{\varsigma}_t) \hat{T}_t + \hat{\mu}^c_{e,t}(\boldsymbol{\varsigma}_t) \hat{f}_t^2(\boldsymbol{\omega}) \nonumber\\
    &+ \hat{\mu}^d_{e,t}(\boldsymbol{\varsigma}_t) \hat{f}_t^4(\boldsymbol{\omega})\label{Temp_CC1}\\
    \mathbb{P}_{\boldsymbol{\varsigma},\boldsymbol{\omega}}[T_t & \leq T^\text{max}] \geq 1-\epsilon,\label{Temp_CC2}
\end{align}
\end{subequations}
where \eqref{Temp_CC1} is the extension of \eqref{new_cons1} under uncertainty. Eq. ~\eqref{Temp_CC2} is the temperature constraint under CC and symbol $\mathbb{P}[\cdot]$ denotes a probability operator. The nonlinearity in both decision and random variables renders the resulting feasibility region non-convex.

\vspace{0.2cm}
\subsection{Approximation of non-convex chance constraints} \label{B.3}
In this paper we model uncertainty via chance constraints, which explicitly incorporate random variables into an optimization problem and ensure that critical operating limits are satisfied with a prescribed probability level. In general, a CC optimization problem has the following form:
\begin{equation}
    \mathbb{P} [q(x,\boldsymbol{\varphi}) \leq 0] \geq 1-\epsilon,
\end{equation}
where $x$ is a decision variable, $\boldsymbol{\varphi}$ represents a random variable, and $\epsilon \in (0,1)$ is a user-specific risk tolerance, and the probability of constraint feasibility must be at least $1-\epsilon$. In power system applications, this is particularly appropriate when operating limits (such as line ratings or reserve margins) are affected by exogenous uncertainties (e.g., weather conditions, or renewable generation) \cite{Group:tomas, bienstock2014chance}. The CC framework is attractive for two main reasons:
\begin{itemize}
    \item Explicit control of operational risk. Unlike robust optimization, which enforces feasibility for all possible realizations in a pre-defined uncertainty set (often leading to overly conservative decisions), CC allows a small, quantified probability of violation. This yields decisions that are significantly less conservative yet remain provably safe with the required probability.
    \item Analytical tractability. For a large class of problems, including affine functions of Gaussian random variables, CC leads to a closed-form deterministic reformulation (e.g., an equivalent second-order cone constraint). This enables the use of efficient convex optimization solvers in practical power system applications \cite{CC:paper1}, \cite{Group:yury}.
\end{itemize}

In this paper, non-convex constraints in \eqref{Temp_CC} increase computational complexity of the OPF. To cope with this, we first consider a general non-convex CC optimization:
\begin{subequations}\label{general}
\begin{align}
    \min_{\hat{x}} & \quad \sum_{t=1}^T C_t(\hat{x}_t) \\
     & \quad \hat{u}_{t+1} = h(\hat{u}_t, \hat{x}_t, \phi_t+\boldsymbol{\varphi}_t), t=1,...,T \label{gen:evolu}\\
    & \quad \mathbb{P}_\varphi [q(\hat{u}_t, \hat{x}_t) \leq 0] \geq 1-\epsilon, t=1,...,T, \label{general:CC}
\end{align}
\end{subequations}
where $\hat{x}_t \in \mathbb{R}^n$ denotes the decision variables, corresponding to generator power outputs $p_t$ in the OPF model. The state variables $\hat{u}_t \in \mathbb{R}^m$ are not directly controllable but evolve according to dynamics in \eqref{gen:evolu}. The uncertainty term $\phi_t+\boldsymbol{\varphi}_t$ corresponds to the wind generation and DLR uncertainty expressions in \eqref{omega_wind} and \eqref{omega_DLR}, with deterministic part $\phi_t$ and addicted stochastic part $\boldsymbol{\varphi}_t$. For analytical tractability, we assume that $q(u,x)$ is non-decreasing.

In the OPF context, $u_t$ corresponds to $f_{e,t}$ and $T_{e,t}$, as described by \eqref{Temp_CC1} and the following equation:
\begin{equation}
    f_{e,t} = \sum_{i\in \mathcal{V}}S_{e,i}(p_{i,t}+w_{i,t}-d_{i,t}),
\end{equation}
where $p_{i,t}$, $d_{i,t}$ and $w_{i,t}$ represent, respectively, the generator power output, load, and forecasted wind generation at node $i$. $S$ is the PTDF matrix for the whole system. 

First, consider a deterministic problem ignoring $\boldsymbol{\varphi}_t$:
\begin{subequations}\label{general:deter}
\begin{align}
    \min_{x} & \quad \sum_{t=1}^T C_t(x_t) \\
    & \quad u_{t+1} = h(u_t, x_t,\phi_t), t=1,...,T \label{deter:evolu}\\
    & \quad q(u_t,x_t) \leq 0, t=1,...,T.
\end{align}
\end{subequations}
Define $\mathbf{X}_{t+1}=[u_{t+1}, u_t, x_t, \phi_t]^\top$ and let  $\mathcal{T}(\mathbf{X}_{t+1}) = 0$ represent \eqref{deter:evolu}. A first-order Taylor approximation at reference point $\breve{\mathbf{X}}_{e,t+1} = [\breve{u}_{t+1},\breve{u}_{t},\breve{x}_{t},\breve{\phi}_{t}]^\top$ yields:
\begin{align}\label{general:TU}
    \mathcal{T}(\mathbf{X}_{t+1}) &= \mathcal{T}(\breve{\mathbf{X}}_{t+1}) + \mathcal{J}(\breve{u}_{t+1})(u_{t+1} - \breve{u}_{t+1}) \nonumber \\
    &+ \mathcal{J}(\breve{u}_{t})(u_{t} - \breve{u}_{t}) + \mathcal{J}(\breve{x}_{t})(x_{t} - \breve{x}_{t})  \\
    &+ \mathcal{J}(\breve{\phi}_{t})(\phi_{t} - \breve{\phi}_{t}) = 0,\nonumber
\end{align}
where $\mathcal{J}(\cdot)$ denotes the Jacobian of $\mathcal{T}$, i.e., partial derivatives relative to its arguments. Since $\mathcal{T}(\breve{\mathbf{X}}_{t+1}) = 0$, \eqref{general:TU} simplifies to:
\begin{equation}
    u_{t+1} = \kappa^0_t + \kappa^u_t u_{t} + \kappa^x_t x_t + \kappa^\phi_t \phi_t, \label{general:kappa1}
\end{equation}
where $\kappa^0_t = \breve{u}_{t+1}+\frac{\mathcal{J}(\breve{u}_{t})}{\mathcal{J}(\breve{u}_{t+1})}\breve{u}_{t}+\frac{\mathcal{J}(\breve{x}_{t})}{\mathcal{J}(\breve{u}_{t+1})}\breve{x}_{t}$, $\kappa^u_t=-\frac{\mathcal{J}(\breve{u}_{t})}{\mathcal{J}(\breve{u}_{t+1})}u_{t}$, $\kappa^x_t=-\frac{\mathcal{J}(\breve{x}_{t})}{\mathcal{J}(\breve{u}_{t+1})}x_{t}$, $\kappa^\phi_t=-\frac{\mathcal{J}(\breve{\phi}_{t})}{\mathcal{J}(\breve{u}_{t+1})}\phi_{t}$.

We now incorporate uncertainty $\boldsymbol{\varphi}_t$ into the equation $h(\cdot)$ in \eqref{deter:evolu}, leading to:
\begin{equation}
    \mathcal{T}(\hat{\mathbf{X}}_{t+1}) = \mathcal{T}(\hat{u}_{t+1}, \hat{u}_t, \hat{x}_t, \phi_t + \boldsymbol{\varphi}_t) =0.
\end{equation}
Applying a first-order Taylor expansion at the same nominal point $\breve{\mathbf{X}}_{e,t+1} = [\breve{u}_{t+1},\breve{u}_{t},\breve{x}_{t},\breve{\phi}_{t}]^\top$, we obtain:
\begin{equation}\label{general:kappa2}
    \hat{u}_{t+1} = \kappa^0_t + \kappa^u_t \hat{u}_{t} + \kappa^x_t \hat{x}_t + \kappa^{\phi}_t (\phi_t + \boldsymbol{\varphi}_t).
\end{equation}
Combining \eqref{general:kappa1} and \eqref{general:kappa2}, the difference between the stochastic and deterministic state becomes:
\begin{equation}
    \hat{u}_{t+1} - u_{t+1} = \kappa^u_t (\hat{u}_{t} - u_{t}) + \kappa^x_t (\hat{x}_{t} - x_{t}) + \kappa^{\phi}_t \boldsymbol{\varphi}_t,
\end{equation}
which indicates that $\hat{u}_{t+1}$ depends on all prior uncertainty realizations $\{\boldsymbol{\varphi}_1, ... ,\boldsymbol{\varphi}_t\}$. To manage this propagation, we introduce non-negative auxiliary reserve variables $R^{\overline{u}}_t$ and $R^{\overline{x}}_t$ as upper bounds on deviations of $\hat{u}_{t} - u_{t}$ and $\hat{x}_{t} - x_{t}$:
\begin{subequations}
    \begin{gather}
    R^{\overline{u}}_t \geq \max\{ \hat{u}_{t} - u_{t}, 0\}\\
    R^{\overline{x}}_t \geq \max\{ \hat{x}_{t} - x_{t}, 0\}\\
    R^{\overline{u}}_{t+1} \geq \kappa^u_t R^{\overline{u}}_t + \kappa^x_t R^{\overline{x}}_t+ \kappa^{\phi}_t \boldsymbol{\varphi}_t.
\end{gather}
\end{subequations}
We thus reformulate the original CC problem \eqref{general} as:
\begin{subequations}\label{general:new}
\begin{align}
    \min_{x} & \quad \sum_{t=1}^T C_t(x_t + R^{\overline{x}}_t) \\
    & \quad u_{t+1} = \kappa^0_t + \kappa^u_t u_{t} + \kappa^x_t x_t + \kappa^{\mu}_t \mu_t\\
    & \quad  R^{\overline{x}}_t \geq 0,  R^{\overline{u}}_t\geq 0,R^{\overline{u}}_1 = 0 \\
    & \quad q(u_{t} + R^{\overline{u}}_t, x_{t} + R^{\overline{x}}_t) \leq 0 \\
    & \quad \mathbb{P}_{\varphi}[R^{\overline{u}}_{t+1} \geq \kappa^u_t R^{\overline{u}}_t + \kappa^x_t R^{\overline{x}}_t+ \kappa^{\phi}_t \boldsymbol{\varphi}_t ] \geq 1-\epsilon ,\label{general:CC_new}
\end{align}
\end{subequations}
where the reformulated CC in \eqref{general:CC_new} is linear and convex, allowing tractable solution methods under various distributional assumptions on $\boldsymbol{\varphi}_t$. In Appendix E, we present a small system to analyze the optimal gap of the proposed relaxation method from both theoretical and simulation perspectives. We then apply the relaxation in \eqref{general:new} to transient DLRs in Section \ref{C.2}.

\vspace{0.2cm}
\section{Pricing with DLRs}\label{C}
\vspace{0.1cm}
\subsection{Singe-period formulation}
We first incorporate the steady-state DLR model in \eqref{DLR_max} into a single-period CC DC-OPF model. Let $\boldsymbol{\omega} = (\boldsymbol{\omega}_i)_{i\in\mathcal{G}} \sim\mathcal{N}(0, \Sigma_\omega)$, $\boldsymbol{\Omega} = \sum_{i\in \mathcal{G}} \boldsymbol{\omega}_i $, $\boldsymbol{\xi_e} \sim \mathcal{N}(0, \sigma_{le})$, the vector $b_{\omega e}$ contains covariances, where $b_{\omega e,k} = \text{Cov}(\boldsymbol{\omega}_k, \boldsymbol{\xi}_e)$. The generator output under uncertainty is modeled using a proportional control law as in \cite{CC:paper1} as $g_i(\boldsymbol{\Omega}) = p_i + \alpha_i \boldsymbol{\Omega}$, where $\alpha_i$ is the participation factor linking the system-wide wind power imbalance $\boldsymbol{\Omega}$ to generator $i$'s adjustment. The single-period CC DC-OPF is formulated as:
\begin{subequations}\label{SCC}
\begin{align}
    \min_{\{p_i,\alpha_i\}} &\quad \mathbb{E}_{\omega,\xi} [\sum_{i\in \mathcal{G}} c_{i,1} (p_i+\alpha_i \boldsymbol{\Omega})+c_{i,2}(p_i+\alpha_i \boldsymbol{\Omega})^2] \\
    (\lambda^\text{bal}):& \sum_{i\in \mathcal{G}}p_i+\sum_{i\in \mathcal{W}}w_i-\sum_{i\in \mathcal{V}}d_i=0 \label{SCC:bal}\\
    (\lambda^\alpha):& \sum_{i\in \mathcal{G}}\alpha_i = 1\label{SCC:sum_a}\\
    (\lambda^{\underline{\alpha}}_i):& \alpha_i \geq 0, \forall i\label{SCC:a+}\\
    (\lambda_{i}^{\overline{p}} ):& p_i + R_i^\text{up} \leq p_i^\text{max}\label{SCC:P_max}\\
    (\lambda_{i}^{\underline{p}}):& p_i - R_i^\text{dn} \geq p_i^\text{min}\label{SCC:P_min}\\
    (\lambda_{i}^{\overline{\text{re}_0}}):& \mathbb{P}_\omega [\alpha_i\boldsymbol{\Omega} \leq R_i^\text{up}] \geq 1-\epsilon ,\forall i \in \mathcal{G}\label{SCC:resmax}\\
    (\lambda_{i}^{\underline{\text{re}_0}}):& \mathbb{P}_\omega [-\alpha_i \boldsymbol{\Omega} \leq R_i^\text{dn}] \geq 1-\epsilon ,\forall i \in \mathcal{G}\label{SCC:resmin}\\
    (\lambda_{e}^{\overline{F_0}} ):& \mathbb{P}_{\omega,\xi} [\sum_{i\in\mathcal{V}} S_{e,i}(p_i+w_i-d_i+\alpha_i \boldsymbol{\Omega}- \boldsymbol{\omega}_i ) \nonumber \\ 
    & \quad\quad \leq f_e^\text{max}+\boldsymbol{\xi}_e]\geq 1-\epsilon,\forall e \in \mathcal{E}\label{SCC:Fmax}\\
    (\lambda_{e}^{\underline{F_0}} ):& \mathbb{P}_{\omega,\xi} [\sum_{i\in\mathcal{V}} S_{e,i}(p_i+w_i-d_i+\alpha_i \boldsymbol{\Omega}- \boldsymbol{\omega}_i )  \nonumber \\
    & \quad\quad \geq -f_e^\text{max}-\boldsymbol{\xi}_e]\geq 1-\epsilon,\forall e \in \mathcal{E},\label{SCC:Fmin}
\end{align}
\end{subequations}
where $R_i^\text{up}$ and $R_i^\text{dn}$ represent the up and down reserves for generator $i$ in response to uncertainty $\boldsymbol{\Omega}$. $\lambda$ is the dual variable corresponding to each constraint. CC \eqref{SCC:resmax} and \eqref{SCC:resmin} enforce generator capacity limits with probability of at least $1-\epsilon$, while CC \eqref{SCC:Fmax} and \eqref{SCC:Fmin} limit the probability of overloaded transmission line to $\epsilon$, with the reserve deliverability of $\alpha_i\boldsymbol{\Omega}$ taken into account.

We use a second-order cone (SOC) reformulation \cite{CC:paper1} to transform the CC-OPF \eqref{SCC} into a deterministic model:
\begin{subequations}\label{SOC}
\begin{align}
    \min_{\{p_i,\alpha_i\}} \quad & \sum_{i\in \mathcal{G}} c_{i,2}(p_i^2+\Sigma_{\Omega} \alpha_i^2) + c_{i,1}p_i & \label{SOC:obj}\\
     \quad & (\text{\ref{SCC:bal}}),(\text{\ref{SCC:sum_a}}),(\text{\ref{SCC:a+}}),(\text{\ref{SCC:P_max}}),(\text{\ref{SCC:P_min}})\nonumber \\
    (\lambda_{i}^{\overline{\text{re}}}): & R_i^\text{up} \geq \Sigma_\Omega^{1/2} \delta \alpha_i, \quad, \forall i \in \mathcal{G}\\
    (\lambda_{i}^{\underline{\text{re}}}): & R_i^\text{dn} \geq \Sigma_\Omega^{1/2} \delta \alpha_i, \quad, \forall i \in \mathcal{G}\\
    (\lambda_{e}^{\overline{F}} ): & \delta \left\| \begin{bmatrix}
        \Sigma_{\omega}^{1/2} (\tilde{S}_e \boldsymbol{\alpha} - \hat{S}_e - \Sigma_{\omega}^{-1} b_{\omega e} ) \\
        \sqrt{ \sigma_{le}^2 - b_{\omega e}^\top \Sigma_{\omega}^{-1} b_{\omega e} }
    \end{bmatrix} \right\|_2 \leq \nonumber \\ 
    & \quad -f_e^\text{max} - \sum_{i\in\mathcal{V}} S_{e,i}(p_i+w_i-d_i)\label{SOC:fmax}\\
    (\lambda_{e}^{\underline{F}} ):& \delta \left\| \begin{bmatrix}
        \Sigma_{\omega}^{1/2} (\tilde{S}_e \boldsymbol{\alpha} - \hat{S}_e - \Sigma_{\omega}^{-1} b_{\omega e} ) \\
        \sqrt{ \sigma_{le}^2 - b_{\omega e}^\top \Sigma_{\omega}^{-1} b_{\omega e} }
    \end{bmatrix} \right\|_2 \leq \nonumber \\ 
    & \quad f_e^\text{max} + \sum_{i\in\mathcal{V}} S_{e,i}(p_i+w_i-d_i),\label{SOC:fmin}
\end{align}
\end{subequations}
where $\delta = \Phi^{-1}(1-\epsilon)$, $\Phi^{-1}(\cdot)$ is the inverse cumulative distribution, $\Sigma_\Omega = \mathbb{1}^\top \Sigma_\omega \mathbb{1}$, $\mathbb{1}$ is a unit vector, $\Tilde{S}_e$ and $\hat{S}_e$ are compact forms of the correlated PTDF matrix components:
\begin{subequations}
\begin{equation}
    \hat{S}_e  = [S_{e,1}, ...,S_{e,|\mathcal{V}|}]^\top \in \mathbb{R}^{|\mathcal{V}|}
\end{equation}
\begin{equation}
    \tilde{S}_e  = [\hat{S}_e, ...,\hat{S}_e]^\top \in \mathbb{R}^{|\mathcal{V}| \times|\mathcal{V}|}.
\end{equation}
\end{subequations}
In \eqref{SOC:fmax} and \eqref{SOC:fmin}, we impose the condition $b_{\omega e}^\top \Sigma_{\omega}^{-1} b_{\omega e} \leq \sigma_{le}^2 $ to guarantee the non-negativity of the expression under the square root, which captures the variance of the DLR over line $e$, with  $\sigma_{le}^2$ being the intrinsic component  and  $b_{\omega e}^\top \Sigma_{\omega}^{-1} b_{\omega e}$ being induced by the correlated wind uncertainty $\boldsymbol{\omega}$. This assumption reflects the physical observation that DLR is more sensitive to local weather variations than to remote ones.

Based on the formulation in \eqref{SOC} and the Lagrangian methods in \cite{CC:paper1}, the LMP and locational marginal reserve price (LMRP) at node $i$ are given by:
\begin{subequations}\label{SOC:LMP}
    \begin{gather}
    \text{LMP}_i = \frac{\partial \mathcal{L}}{\partial d_i} = \lambda^\text{bal} - \sum_{e\in \mathcal{E}} {\left[ (\lambda_e^{\overline{F}} - \lambda_e^{\underline{F}}) S_{e,i}\right]}\label{LMP:CC}\\
    \text{LMRP}_i = \frac{\partial \mathcal{L}}{\partial(R_i^\text{up} + R_i^\text{dn})} =\lambda_i^{\overline{\text{re}}} + \lambda_i^{\underline{\text{re}}}, \label{LMRP:CC}
    \end{gather}
\end{subequations}
where $\mathcal{L}$ denotes the Lagrangian function of (\ref{SOC}). We calculate LMPs by differentiating $\mathcal{L}$ to local demand $d_i$, and calculate LMRPs from the derivative of $\mathcal{L}$ to reserves $R_i^\text{up} + R_i^\text{dn}$. 

These expressions follow directly from partial KKT conditions. The stationary conditions for $p_i$, $\alpha_i$, $R_i^\text{up}$ and $R_i^\text{dn}$ are:
\begin{subequations}
\begin{align}
    \frac{\partial \mathcal{L}}{\partial p_i}: & c_{i,1}+2c_{i,2}p_i + \lambda_{i}^{\overline{p}} - \lambda_{i}^{\underline{p}} - \lambda^\text{bal} + \nonumber\\
    & \sum_{e\in \mathcal{E}}{\left[ (\lambda_e^{\overline{F}} - \lambda_e^{\underline{F}}) S_{e,i}\right]}=0 \label{KKT-CC:sys:p}\\
    \frac{\partial \mathcal{L}}{\partial \alpha_i}: & \sum_{e\in \mathcal{E}} (\lambda_e^{\overline{F}} - \lambda_e^{\underline{F}})Q^s_{e}\delta + 2c_{i,2}\Sigma_{\Omega}\alpha_i + \nonumber\\
    & (\lambda_i^{\overline{\text{re}}} + \lambda_i^{\underline{\text{re}}}) \Sigma_\Omega^{1/2} \delta -\lambda^\alpha =0 \label{KKT-CC:sys:a}\\
    \frac{\partial \mathcal{L}}{\partial R_i^\text{up}}: & \lambda_{i}^{\overline{p}} - \lambda_i^{\overline{\text{re}}} = 0\\
    \frac{\partial \mathcal{L}}{\partial R_i^\text{dn}}: & \lambda_{i}^{\underline{p}} - \lambda_i^{\underline{\text{re}}} = 0,
\end{align}
\end{subequations}
where
\begin{equation}
    Q^s_{e}= \frac{ {\Sigma}_{\omega} (\tilde{S}_e \boldsymbol{\alpha} - \hat{S}_e - \Sigma_{\omega}^{-1} b_{\omega e} )}{\left\| \begin{bmatrix}
        \tilde{S}_e^\top \Sigma_{\omega}^{1/2} (\tilde{S}_e \boldsymbol{\alpha} - \hat{S}_e - \Sigma_{\omega}^{-1} b_{\omega e} ) \\
        \sqrt{ \sigma_{le}^2 - b_{\omega e}^\top \Sigma_{\omega}^{-1} b_{\omega e} }
    \end{bmatrix} \right\|_2}.
\end{equation}
Then the LMRPs can also be represented as:
\begin{align}\label{SOC:LMRP}
    \text{LMRP}_i \!=\! 
    \frac{1}{\Sigma_\Omega^{1/2} \delta} [ \lambda_i^\alpha \!-\! 2c_{i,2}\Sigma_\Omega\alpha_i \!-\! \sum_{e\in \mathcal{E}} (\lambda_e^{\overline{F}} \!-\! \lambda_e^{\underline{F}})Q^s_{e}\delta ],
\end{align}
where $\lambda_i^\alpha$ is reserve balance price, $2c_{i,2}\Sigma_\Omega\alpha_i$ captures the local generator reserve cost, and $\sum_{e\in \mathcal{E}} (\lambda_e^{\overline{F}} - \lambda_e^{\underline{F}})Q^s_{e}\delta$ reflects reserve delivery cost.

We now show that the LMP and LMRP formulations in \eqref{LMP:CC} and \eqref{SOC:LMRP} lead to a market equilibrium.

\begin{theorem}[Market equilibrium for single-period OPF]\label{theorem:market_single}
Let $\{ p_i^*, \alpha_i^*\}$ be the optimal solution of the problem in \eqref{SOC} and let $\{\pi_i^*, \tau_i^*\}$ be the LMP and LMRP calculated by \eqref{LMP:CC} and \eqref{SOC:LMRP} respectively. Then $\{p_i^*, \alpha_i^*, \pi_i^*, \tau_i^*\}$ constitutes a market equilibrium, i.e.:
\begin{itemize}
\item The market clears at $\sum_{i\in \mathcal{G}}p_i+\sum_{i\in \mathcal{W}}w_i-\sum_{i\in \mathcal{V}}d_i=0 $ and $\sum_{i\in \mathcal{G}}\alpha_i = 1$
\item Each producer maximizes its profit under payment $\Gamma_i = \pi^*_i p_i^* + \tau_i^* \alpha_i^*$
\hfill \qed
\end{itemize}
\end{theorem}

The proof is given in Appendix \ref{app:market_single}. In this equilibrium, the LMPs and LMRPs, as determined by \eqref{LMP:CC} and \eqref{SOC:LMRP}, provide the correct economic signals such that each generator, behaving as a price taker in a perfectly competitive market, maximizes its individual profit, while the system-level objective of social welfare maximization, as formulated in \eqref{SOC}, is simultaneously achieved.

\subsection{Multi-period formulation}\label{C.2}

We extend \eqref{SCC} to multiple time periods to  incorporate the transient DLRs and account for the correlated uncertainty of wind generation and transient DLRs as in \eqref{omega_wind} and \eqref{omega_DLR}. The multi-period CC DC-OPF is:
\begin{subequations}\label{MU-OPF}
\begin{align}
    \min \quad& \mathbb{E}_{\omega,\varsigma}[\sum_{t}\sum_{i\in \mathcal{G}} c_{i,1} (p_{i,t}+\alpha_{i,t}\Omega_{i,t})\nonumber\\
    & \quad + c_{i,2}(p_{i,t}+\alpha_{i,t}\Omega_{i,t})^2] \label{MU-OPF:obj}\\
    (\lambda_t^\text{bal}):& \sum_{i\in \mathcal{G}}p_{i,t}+\sum_{i\in \mathcal{W}} w_{i,t}-\sum_{i\in \mathcal{V}}d_{i,t}=0, \forall t \label{MU-OPF:bal}\\
    (\lambda_t^\alpha):& \sum_{i\in \mathcal{G}}\alpha_{i,t} = 1, \forall t \label{MU-OPF:alpha}\\
    (\lambda^{\underline{\alpha}}_{i,t}):& \alpha_{i,t} \geq 0, \forall i\in \mathcal{G}, \forall t \label{MU-OPF:a+}\\
    (\lambda_{i,t}^{\overline{p}}):& p_{i,t} + R_{i,t}^\text{up} \leq p_i^\text{max},\forall i \in \mathcal{G}, \forall t \label{MU-OPF:gmax}\\
    (\lambda_{i,t}^{\underline{p}}):& p_{i,t} - R_{i,t}^\text{dn} \leq p_i^\text{min},\forall i \in \mathcal{G}, \forall t \label{MU-OPF:gmin}\\
    (\lambda_{i,t}^{\underline{\text{rr}}}):&  p_{i,t+1} - p_{i,t} + R^\text{up}_{i,t+1} + R^\text{dn}_{i,t} \leq U_i^\text{up} , \forall i , \forall t \label{MU-OPF:UP}\\
    (\lambda_{i,t}^{\underline{\text{rr}}}):& p_{i,t+1} - p_{i,t} - R^\text{dn}_{i,t+1} - R^\text{up}_{i,t} \geq -U_i^\text{dn} ,\forall i , \forall t \label{MU-OPF:DN}\\
    (\lambda_{i,t}^{\overline{\text{re}_0}}):& \mathbb{P}_\omega [R_{i,t}^\text{up} \geq \alpha_{i,t}\Omega_{i,t}] \geq 1 - \epsilon,\forall i \in \mathcal{G}, \forall t \label{MU-OPF:rev1}\\
    (\lambda_{i,t}^{\underline{\text{re}_0}}):& \mathbb{P}_\omega [R_{i,t}^\text{dn} \geq -\alpha_{i,t}\Omega_{i,t}] \geq 1-\epsilon,\forall i \in \mathcal{G}, \forall t \label{MU-OPF:rev2}\\
    (\lambda_{e,t}^{f'}):& \hat{f}_{e,t}(\boldsymbol{\omega}_t) = \sum_{i\in\mathcal{V}} S_{e,i}(p_{i,t}+w_{i,t}-d_{i,t}+\nonumber\\
    &\quad \quad \alpha_{i,t} \Omega_{i,t}-\boldsymbol{\omega}_{i,t} ),\forall e \in \mathcal{E},\forall t \label{MU-OPF:f}\\
    (\lambda_{e,t}^{T'}):& \hat{T}_{e,t+1}(\boldsymbol{\omega}_t,\boldsymbol{\varsigma}_{e,t}) = \hat{\mu}^a_{e,t}(\boldsymbol{\varsigma}_{e,t}) \hat{T}_{e,t}(\boldsymbol{\omega}_{t-1},\boldsymbol{\varsigma}_{e,t-1}) +\nonumber\\
    &\hat{\mu}^b_{e,t}(\boldsymbol{\varsigma}_{e,t}) \hat{f}^2_{e,t}(\boldsymbol{\omega}_t) + \hat{\mu}^c_{e,t}(\boldsymbol{\varsigma}_{e,t}) \hat{f}_{e,t}^4(\boldsymbol{\omega}_t),\forall e, \forall t \label{MU-OPF:T}\\
    (\lambda_{e,t}^{\overline{T'}}):&\mathbb{P}_{\omega,\varsigma} [ \hat{T}_{e,t}(\boldsymbol{\omega}_t,\boldsymbol{\varsigma}_{e,t}) \leq T^\text{max}_e] \geq 1-\epsilon,\forall e, \forall t, \label{MU-OPF:Tmax}
\end{align}
\end{subequations}
where \eqref{MU-OPF:T} models the temperature evolution process, capturing the combined effects of wind power and DLR uncertainties. As is discussed in Section II, \eqref{MU-OPF:T} and \eqref{MU-OPF:Tmax} introduce non-convexity. We approximate this CC via linearization and first consider its deterministic form:
\begin{equation}\label{evo_fun}
    T_{e,t+1} = \mu_{e,t}^a + \mu_{e,t}^b T_{e,t} + \mu_{e,t}^c f_{e,t}^2 + \mu_{e,t}^d f_{e,t}^4.
\end{equation}
We denote \eqref{evo_fun} as $\mathcal{T}(\mathbf{T}_{e,t+1})=0$ with $\mathbf{T}_{e,t+1} = [T_{e,t+1},T_{e,t},f_{e,t},\mu_{e,t}^a,\mu_{e,t}^b,\mu_{e,t}^c,\mu_{e,t}^d]^\top$. Then, we apply a first-order Taylor expansion of $\mathcal{T}(\cdot)=0$ at the reference point $\breve{\mathbf{T}}_{e,t+1} = [\breve{T}_{e,t+1},\breve{T}_{e,t},\breve{f}_{e,t},\breve{\mu}_{e,t}^a,\breve{\mu}_{e,t}^b,\breve{\mu}_{e,t}^c,\breve{\mu}_{e,t}^d]^\top$, yielding:
\begin{equation}\label{big_T}
    \!\!\!\!\mathcal{T}(\mathbf{T}_{e,t+1})\!\!=\!\! \mathcal{T}(\breve{\mathbf{T}}_{e,t+1})\! +\! \mathcal{J}_{\mathcal{T}}(\breve{\mathbf{T}}_{e,t+1})^\top\!(\mathbf{T}_{e,t+1}\!\! -\!\! \breve{\mathbf{T}}_{e,t+1})\!\! =\!\! 0,
\end{equation}
where $\mathcal{J}_{\mathcal{T}}$ is the Jacobian of $\mathcal{T}$. Let $\mathbf{U}_{e,t}=[T_{e,t},f_{e,t},\mu_{e,t}^a,\mu_{e,t}^b,\mu_{e,t}^c,\mu_{e,t}^d]^\top$, $\kappa_{e,t+1} = \frac{\mathcal{J}_{\mathcal{T}}(\breve{\mathbf{U}}_{e,t})}{\mathcal{J}_{\mathcal{T}}(\breve{T}_{e,t+1})}$, $\kappa^a_{e,t+1} = \mathcal{J}_{\mathcal{T}}(\breve{T}_{e,t+1})$. Since $\mathcal{T}(\breve{\mathbf{T}}_{e,t+1}) = 0$, \eqref{big_T} can be:
\begin{equation}
    T_{e,t+1} =\kappa^a_{e,t+1} + \kappa_{e,t+1}^\top \mathbf{U}_{e,t}.\label{Linear}
\end{equation}
Using \eqref{Linear}, \eqref{MU-OPF:T} can be linearized as: 
\begin{table*}[!b]
    \centering
    \caption{Comparison of Pricing Outcomes under Different Line Rating Models}
    \begin{tabular}{|c|c|c|}
        \hline
        \textbf{Line rating models}& \textbf{Single-period OPF} & \textbf{Multi-period OPF} \\
        \hline
        \textbf{SLR} & $\text{LMP}_{i} = \lambda^\text{bal} - \sum_{e\in \mathcal{E}} S_{e,i}\lambda_{e}^f$ & $\text{LMP}_{i,t} = \lambda_t^\text{bal} - \sum_{e\in \mathcal{E}} S_{e,i}\lambda_{e,t}^f$ \\

        \hline
        \textbf{DLR} & $\text{LMP}_{i} = \lambda^\text{bal} - \sum_{e\in \mathcal{E}} S_{e,i}\lambda_{e}^f$ & $\text{LMP}_{i,t} = \lambda_t^\text{bal} - \sum_{e\in \mathcal{E}} S_{e,i}\lambda_{e,t}^f$ \\
        \hline
        \textbf{Stochastic DLR} & $\text{LMP}_{i} = \lambda^\text{bal} - \sum_{e\in \mathcal{E}} S_{e,i}\lambda_{e}^f$ & $\text{LMP}_{i,t} = \lambda_t^\text{bal} - \sum_{e\in \mathcal{E}} S_{e,i}\lambda_{e,t}^f$ \\
        & $\text{LMRP}_i = \frac{\lambda^\alpha_i}{\Omega} - 2 c_{i,2}\Omega \alpha_i - \sum_{e\in \mathcal{E}}S_{e,i}(\lambda_e^{\overline{F}} - \lambda_e^{\underline{F}})$ & $\text{LMRP}_{i,t} = \frac{\lambda_t^\alpha}{\Omega_t} - 2 c_{i,2}\Omega_t \alpha_{i,t} + \sum_{e\in \mathcal{E}}S_{e,i}\lambda_{e,t}^f $ \\
        \hline
         & $\text{LMP}_{i} = \lambda^\text{bal} - \sum_{e\in \mathcal{E}} S_{e,i}\lambda_{e}^f$ & $\text{LMP}_{i,t} = \lambda_t^\text{bal} -\frac{1}{2}\sum_{e\in\mathcal{E}}S_{e,i} \times $ \\ 
        \textbf{CC DLR} & $\text{LMRP}_i = \frac{1}{\Sigma_\Omega^{1/2} \delta} [ \lambda_i^\alpha - 2c_{i,2}\Sigma_\Omega \alpha_i - \sum_{e\in \mathcal{E}} (\lambda_e^{\overline{F}} - \lambda_e^{\underline{F}})Q^s_{e,t}\delta ]$ & $\left[\kappa_{e,t+1}^b \lambda_{e,t+1}^T \!-\! \lambda_{e,t+1}^{\overline{T}} \!+\! \frac{\lambda_{e,t-1}^T \!+ \! \lambda_{e,t}^{\overline{T}}}{\kappa_{e,t}^b}\right]$\\
        &  & $\text{LMRP}_{i,t} = \frac{1}{\Sigma_\Omega^{1/2} \delta} [\lambda^\alpha_t - 2 c_{i,2} \Sigma_\Omega \alpha_{i,t} - \sum_{e\in \mathcal{E}}\lambda_{e,t}^{\overline{\text{th}}}  Q^m_{e,t} \delta ]$ \\
        \hline
    \end{tabular}
    \label{tab:comparison}
\end{table*}

\begin{gather}
    \hat{T}_{e,t+1} = \kappa^a_{e,t+1}+ \kappa_{e,t+1}^\top \mathbf{\hat{U}}_{e,t}(\boldsymbol{\omega}_t,\boldsymbol{\varsigma}_{e,t}),\label{hatT1}
\end{gather}
where $\mathbf{\hat{U}}_{e,t}=[\hat{T}_{e,t},\hat{f}_{e,t},\hat{\mu}_{e,t}^a,\hat{\mu}_{e,t}^b,\hat{\mu}_{e,t}^c,\hat{\mu}_{e,t}^d]^\top$.

According to \eqref{MU-OPF:f}, the stochastic variable $\hat{f}_{e,t}(\boldsymbol{\omega}_t)$ can be reformulated according to deterministic value $f_{e,t}$ as: 
\begin{align}\label{hatf}
    \hat{f}_{e,t}(\boldsymbol{\omega}_t) =& f_{e,t} + \Sigma_{i \in \mathcal{V}} S_{e,i}(\alpha_{i,t}\Sigma_{j\in \mathcal{W}} \boldsymbol{\omega}_{j,t} - \boldsymbol{\omega}_{i,t}) \nonumber \\
    =& f_{e,t} + \acute{\kappa}_{e,t} \alpha_t \boldsymbol{\omega}_t,
\end{align}
where $\acute{\kappa}_{e,t}$ is a matrix to transform $\Sigma_{i \in \mathcal{V}} S_{e,i}(\alpha_{i,t}\Sigma_{j\in \mathcal{W}} \boldsymbol{\omega}_{j,t} - \boldsymbol{\omega}_{i,t})$ into $\acute{\kappa}_{e,t} \alpha_t \boldsymbol{\omega}_t$ for simplicity.

Denote $\boldsymbol{\mu}_{e,t}=[\mu_{e,t}^a,\mu_{e,t}^b,\mu_{e,t}^c,\mu_{e,t}^d]^\top$, $\kappa^\mu_{e,t+1} = \frac{\mathcal{J}_{\mathcal{T}}(\breve{\boldsymbol{\mu}}_{e,t})}{\mathcal{J}_{\mathcal{T}}(\breve{T}_{e,t+1})}$, $\kappa^b_{e,t+1} = \frac{\mathcal{J}_{\mathcal{T}}(\breve{T}_{e,t})}{\mathcal{J}_{\mathcal{T}}(\breve{T}_{e,t+1})}$, $\kappa^c_{e,t+1} = \frac{\mathcal{J}_{\mathcal{T}}(\breve{f}_{e,t})}{\mathcal{J}_{\mathcal{T}}(\breve{T}_{e,t+1})}$. Combine \eqref{omega_DLR} and \eqref{hatf} with \eqref{hatT1}, we can get:
\begin{align}\label{thermal_relax}
    \hat{T}_{e,t+1} - T_{e,t+1}  =& \kappa^b_{e,t+1} (\hat{T}_{e,t} - T_{e,t})  + \kappa^c_{e,t+1} \acute{\kappa}_e \alpha_t \boldsymbol{\omega}_t \nonumber \\
    &+ \kappa_{e,t+1}^{\mu\top}\gamma_{e,t} \boldsymbol{\varsigma}_{e,t} \nonumber \\
    =& \kappa^b (\hat{T}_{e,t} -  T_{e,t}) + \acute{\kappa}^c \alpha_t \boldsymbol{\omega}_t + \acute{\kappa}^d \boldsymbol{\varsigma}_{e,t},
\end{align}
where $\gamma_{e,t} = \text{diag}\{ \gamma_{e,t}^a, \gamma_{e,t}^b,\gamma_{e,t}^c, \gamma_{e,t}^e \}$, $\acute{\kappa}^c = \kappa^c_{e,t+1} \acute{\kappa}_e$, $\acute{\kappa}^d=\kappa_{e,t+1}^{\mu\top}\gamma_{e,t}$. For brevity, we suppress subscripts and write $\acute{\kappa}$ instead of $\acute{\kappa}_{e,t+1}$. Unless otherwise specified, all $\kappa$ and $\acute{\kappa}$ remain line- and time-specific. Since $\hat{T}_{e,t+1}$ depends on $\{ \boldsymbol{\omega}_1,...,\boldsymbol{\omega}_t,\boldsymbol{\varsigma}_1,...,\boldsymbol{\varsigma}_t \}$, we introduce the auxiliary thermal reserve $R^\text{th}_{e,t}$ to relax $\hat{T}_{e,t} - T_{e,t}$:
\begin{align}
    R^\text{th}_{e,t+1} &= \max_{\boldsymbol{\omega},\boldsymbol{\varsigma}} \left\{\hat{T}_{e,t+1} - T_{e,t+1}\right\} \nonumber\\
    &=\max_{\boldsymbol{\omega},\boldsymbol{\varsigma}} \left\{\kappa^b (\hat{T}_{e,t} -  T_{e,t}) + \acute{\kappa}^c \alpha_t \boldsymbol{\omega}_t + \acute{\kappa}^d \boldsymbol{\varsigma}_{e,t}\right\} \nonumber\\
    &= \kappa^b R^\text{th}_{e,t} + \acute{\kappa}^c \alpha_t \boldsymbol{\omega}_t + \acute{\kappa}^d \boldsymbol{\varsigma}_{e,t}.
\end{align}
Note that this auxiliary thermal reserve $R^\text{th}_{e,t}$ represents the spare transmission thermal capacity and is not related  to the primary/secondary/tertiary reserve products for frequency and contingencies. This thermal reserve is introduced to solve the combined uncertainty and non-convexity. 
Then \eqref{thermal_relax} can be relaxed as:
\begin{subequations}\label{41}
    \begin{gather}
        R^{\text{th}}_{e,t} \geq \hat{T}_{e,t} -  T_{e,t} \\
        \begin{cases}
            R^\text{th}_{e,1} = 0 \\
            R^\text{th}_{e,t+1} \geq \kappa^b R^\text{th}_{e,t} + \acute{\kappa}^c \alpha_t \boldsymbol{\omega}_t + \acute{\kappa}^d \boldsymbol{\varsigma}_{e,t}, \text{ for }  t\geq 1.
        \end{cases}
    \end{gather}
\end{subequations}
Then by incorporating \eqref{Linear} and \eqref{41}, the initial constraints \eqref{MU-OPF:f} - \eqref{MU-OPF:Tmax} can be replaced by:
\begin{subequations}
    \begin{align}
        (\lambda_{e,t}^f):& f_{e,t} =  \sum_{i\in\mathcal{V}} S_{e,i}(p_{i,t}+w_{i,t}-d_{i,t}) \label{MC-OPF:f} \\
        (\lambda_{e,t}^T):& T_{e,t+1} = \kappa^a + \kappa^\top[T_{e,t},f_{e,t},\mu_{e,t}^a,\mu_{e,t}^b,\mu_{e,t}^c,\mu_{e,t}^d] \label{MC-OPF:T_evo}\\
                & R^\text{th}_{e,1} = 0 \label{MC-OPF:R_th_1} \\
        (\lambda_{e,t}^{\overline{T}}):& T_{e,t} + R^\text{th}_{e,t} \leq T^\text{max}_e  \label{MC-OPF:T_max}\\
        & \mathbb{P}_{\omega,\varsigma} [ R^\text{th}_{e,t+1} \geq \kappa^b R^\text{th}_{e,t} + \acute{\kappa}^c \alpha_t \boldsymbol{\omega}_t + \acute{\kappa}^d \boldsymbol{\varsigma}_{e,t}] \nonumber \\
        & \quad \quad \quad \quad \quad \quad \geq 1-\epsilon,\forall e \in \mathcal{E}, \forall t.
    \end{align}
\end{subequations}

Assume $\boldsymbol{\varsigma} \sim \mathcal{N}(0, \Sigma_{\varsigma})$, the vector $b_{\omega e}$ contains covariances, where $b_{\omega e,k} = \text{Cov}(\boldsymbol{\omega}_k, \boldsymbol{\varsigma}_e)$. Then all CCs can be transformed into SOC constraints:
\begin{subequations}
    \begin{align}
        (\lambda_{e,t}^{\overline{\text{re}}}):&  \Sigma_\Omega^{1/2} \delta \alpha_i \leq R^\text{up}_{i,t}, \forall i \in \mathcal{G}, \forall t \label{MC-OPF:R_up}\\
        (\lambda_{e,t}^{\underline{\text{re}}}):&  \Sigma_\Omega^{1/2} \delta \alpha_i \leq R^\text{dn}_{i,t} , \forall i \in \mathcal{G}, \forall t \label{MC-OPF:R_dn}\\
        (\lambda_{e,t}^{\overline{\text{th}}}):& \delta \left\| \begin{bmatrix}
        \Sigma_{\omega}^{1/2} \acute{\kappa}^c \alpha_t + \Sigma_{\omega}^{-1/2} \acute{\kappa}^d b_{\omega e}^\top  \\
        \sqrt{\acute{\kappa}^{d\top} \Sigma_\varsigma \acute{\kappa}^d - b_{\omega e} \acute{\kappa}^{d\top} \Sigma_{\omega}^{-1} \acute{\kappa}^d b_{\omega e}^\top}
    \end{bmatrix} \right\|_2 \leq \nonumber \\
    &\quad R^\text{th}_{e,t+1} - \kappa^b R^\text{th}_{e,t}, \forall e \in \mathcal{E}, \forall t, \label{MC-OPF:R_th}
    \end{align}
\end{subequations}
where $\delta = \Phi^{-1}(1-\epsilon)$. In \eqref{MC-OPF:R_th} ,we impose the condition $b_{\omega e} \acute{\kappa}^{d\top} \Sigma_{\omega}^{-1} \acute{\kappa}^d b_{\omega e}^\top \leq \acute{\kappa}^{d\top} \Sigma_\varsigma \acute{\kappa}^d$ to ensure the non-negativity of the expression under the square root, which captures the variance of $\text{DLR}_e$, where $\sigma_{\varsigma}^2$ denotes its intrinsic variance, and $b_{\omega e} \acute{\kappa}^{d\top} \Sigma_{\omega}^{-1} \acute{\kappa}^d b_{\omega e}^\top$ quantifies the contribution from correlated wind uncertainty $\boldsymbol{\omega}$. This assumption reflects the physical observation that DLR is more sensitive to local weather conditions rather than to remote ones.

The initial multi-period CC DC-OPF problem in \eqref{MU-OPF} can be reformulated as:
\begin{align}\label{MC-OPF}
    \min_\mathcal{P} &\quad \sum_{t}\sum_{i\in \mathcal{G}} c_{i,2}(p_{i,t}^2+\Sigma_{\Omega} \alpha_{i,t}^2) + c_{i,1}p_{i,t} \\
    &\quad (\text{\ref{MU-OPF:bal}}), (\text{\ref{MU-OPF:alpha}}), (\text{\ref{MU-OPF:a+}}), (\text{\ref{MU-OPF:gmax}}), (\text{\ref{MU-OPF:gmin}}), (\text{\ref{MU-OPF:UP}}), (\text{\ref{MU-OPF:DN}}), \nonumber\\
    & \quad(\text{\ref{MC-OPF:f}}), (\text{\ref{MC-OPF:T_evo}}), (\text{\ref{MC-OPF:R_th_1}}), (\text{\ref{MC-OPF:T_max}}), (\text{\ref{MC-OPF:R_up}}), (\text{\ref{MC-OPF:R_dn}}), (\text{\ref{MC-OPF:R_th}}) \nonumber,
\end{align}
where $\mathcal{P} = \{ p_{i,t},\alpha_{i,t},f_{e,t},T_{e,t},R^\text{up}_{i,t}, R^\text{dn}_{i,t}, R^\text{th}_{e,t}  \}$, and then the reformulated problem \eqref{MC-OPF} is convex.

Using \eqref{MC-OPF}, we obtain the LMP and LMRP at node $i$:
\begin{subequations}\label{MC-OPF:LMP}
    \begin{equation}
        \text{LMP}_{i,t} = \frac{\partial\mathcal{L}}{\partial d_{i,t}}=\lambda_t^\text{bal} - \sum_{e\in \mathcal{E}} S_{e,i}\lambda_{e,t}^f \label{MC-OPF:LMPa}
    \end{equation}
    \begin{equation}
        \text{LMRP}_{i,t} = \frac{\partial \mathcal{L}}{\partial (R_{i,t}^\text{up} + R_{i,t}^\text{dn})} = {\lambda_{i,t}^{\overline{\text{re}}}} + \lambda_{i,t}^{\underline{\text{re}}}, \label{MC-OPF:LMPb}
    \end{equation}
\end{subequations}
which can be derived due to partial KKT conditions:
\begin{subequations}\label{Multi_KKT}
\begin{align}
    (p_{i,t}):& 2c_{i,2} p_{i,t} + c_{i,1} - \lambda^\text{bal}_t + \lambda_{i,t}^{\overline{p}} - \lambda_{i,t}^{\underline{p}} - \lambda_{i,t}^{\overline{\text{rr}}} + \lambda_{i,t}^{\underline{\text{rr}}} \nonumber\\
    &+ \lambda_{i,t-1}^{\overline{\text{rr}}} - \lambda_{i,t-1}^{\underline{\text{rr}}} + \sum_{e\in \mathcal{E}}S_{e,i}\lambda_{e,t}^f = 0 \label{Multi-KKT:p}\\
    (\alpha_{i,t}):&  2 c_{i,2} \Sigma_\Omega \alpha_{i,t} -\lambda^\alpha_t + ({\lambda_{i,t}^{\overline{\text{re}}}} + \lambda_{i,t}^{\underline{\text{re}}})\Sigma_\Omega^{1/2} \delta \nonumber\\
    &+ \sum_{e\in \mathcal{E}}\lambda_{e,t}^{\overline{\text{th}}}  Q^m_{e,t} \delta = 0\label{Multi-KKT:a} \\
    (f_{e,t}):& \lambda_{e,t}^f - \kappa_{e,t}^c \lambda_{e,t}^T = 0\\
    (T_{e,t}):& \lambda_{e,t-1}^T - \kappa^b_{e,t}+\lambda_{e,t}^{\overline{T}}=0,
\end{align}
\end{subequations}
where $Q_{e,t}^m = \frac{\acute{\kappa}^{c\top} \Sigma_{\omega} \acute{\kappa}^c \alpha_t + \Sigma_{\omega}^{-1} \acute{\kappa}^d b_{\omega e}^\top}{\left\| \begin{bmatrix}
        \Sigma_{\omega}^{1/2} \acute{\kappa}^c \alpha_t + \Sigma_{\omega}^{-1/2} \acute{\kappa}^d b_{\omega e}^\top  \\
        \sqrt{\acute{\kappa}^{d\top} \Sigma_\varsigma \acute{\kappa}^d - b_{\omega e} \acute{\kappa}^{d\top} \Sigma_{\omega}^{-1} \acute{\kappa}^d b_{\omega e}^\top}
    \end{bmatrix} \right\|_2}$.
Then we can reformulate the LMPs and LMRPs as:
\begin{subequations}
    \begin{align}
    \text{LMP}_{i,t} =& \lambda_t^\text{bal} -\frac{1}{2}\sum_{e\in\mathcal{E}}S_{e,i} \times \nonumber\\
    &\left[\kappa_{e,t+1}^b \lambda_{e,t+1}^T \!-\! \lambda_{e,t+1}^{\overline{T}} \!+\! \frac{\lambda_{e,t-1}^T \!+ \! \lambda_{e,t}^{\overline{T}}}{\kappa_{e,t}^b}\right]
    \end{align}
    \begin{equation}
    \text{LMRP}_{i,t} = \frac{1}{\Sigma_\Omega^{1/2} \delta} [\lambda^\alpha_t - 2 c_{i,2} \Sigma_\Omega \alpha_{i,t} - \sum_{e\in \mathcal{E}}\lambda_{e,t}^{\overline{\text{th}}}  Q_{e,t} \delta ], \label{MC-OPF:LMRP}
    \end{equation}
\end{subequations}
which shows that the LMPs at time $t$ are affected by the thermal state and power flow at both time $t-1$ and time $t+1$, indicating that LMPs are influenced by the time-varient interdependency introduced by transient DLRs. In LMRPs, $\lambda_i^\alpha$ is reserve balance price, $2c_{i,2} \Sigma_\Omega\alpha_i$ is local generator reserve cost, and $\sum_{e\in \mathcal{E}}\lambda_{e,t}^{\overline{\text{th}}}  Q_{e,t} \delta$ is reserve delivery cost.

We now show that the LMP and LMRP formulations in \eqref{MC-OPF:LMPa} and \eqref{MC-OPF:LMRP} naturally induce a market equilibrium.

\begin{theorem}[Market equilibrium for multi-period OPF]\label{theorem:market_multi}
Let $\{ p_{i,t}^*, \alpha_{i,t}^*, f_{e,t}^*, T_{e,t}^*, R_{i,t}^{\text{up}*}, R_{i,t}^{\text{dn}*},R_{i,t}^{\text{th}*}\}$ be the optimal solution of the problem in \eqref{MC-OPF} and let $\{\pi_{i,t}^*, \tau_{i,t}^*\}$ be the LMP and LMRP calculated by \eqref{MC-OPF:LMPa} and \eqref{MC-OPF:LMRP} respectively. Then $\{ p_{i,t}^*, \alpha_{i,t}^*, f_{e,t}^*, T_{e,t}^*, R_{i,t}^{\text{up}*}, R_{i,t}^{\text{dn}*},R_{i,t}^{\text{th}*}, \pi_i^*, \tau_i^*\}$ constitutes a market equilibrium, i.e.:
\begin{itemize}
\item The market clears at $\sum_{i\in \mathcal{G}}p_{i,t}+\sum_{i\in \mathcal{W}}w_{i,t}-\sum_{i\in \mathcal{V}}d_{i,t}=0 $ and $\sum_{i\in \mathcal{G}}\alpha_{i,t} = 1$ for $\forall t$
\item Each producer maximizes its profit under the payment $\Gamma_{i} = \sum_{t} (\pi^*_{i,t} p_{i,t}^* + \tau_{i,t}^* \alpha_{i,t}^*$)
\hfill \qed

\end{itemize}
\end{theorem}

The proof is given in Appendix \ref{app:market_multi}. Theorem \ref{theorem:market_multi} implies that, given the LMPs and LMRPs computed from \eqref{MC-OPF:LMPa} and \eqref{MC-OPF:LMRP}, each generator, acting as a price taker, maximize its individual profit while the overall social welfare is maximized as formulated in \eqref{MC-OPF} despite relaxations.

\subsection{Comparison of LMP and LMRP under different line ratings}
Table \ref{tab:comparison} summarizes the formulation of LMPs and LMRPs under different line rating models (SLR, DLR, and CC DLR). We adopt steady-state and transient DLR models in the single- and multi-period OPF, respectively. In both the single- and multi-period settings, the LMP consists of total energy balance price and congestion price characterized by $\sum_{e\in\mathcal{E}}S_{e,i}\lambda^f_e$. The LMRP reflects the total system reserve balance cost, local generator reserve cost, and reserve delivery cost, which differs depending on the adopted DLR model. For steady-state DLR models in single-period problem, it is expressed using the direct power flow limit as $\sum_{e\in \mathcal{E}} (\lambda_e^{\overline{F}} - \lambda_e^{\underline{F}})Q^s_{e,t}\delta$, while for transient thermal models, it is expresed in terms of thermal constraints as $\sum_{e\in \mathcal{E}}\lambda_{e,t}^{\overline{\text{th}}}  Q^m_{e,t} \delta$.

\vspace{0.5cm}
\section{Numerical Experiments}\label{E}

\subsection{Approximation of line temperature evolution}
\begin{figure}[t]
    \centering
    \includegraphics[width=\columnwidth]{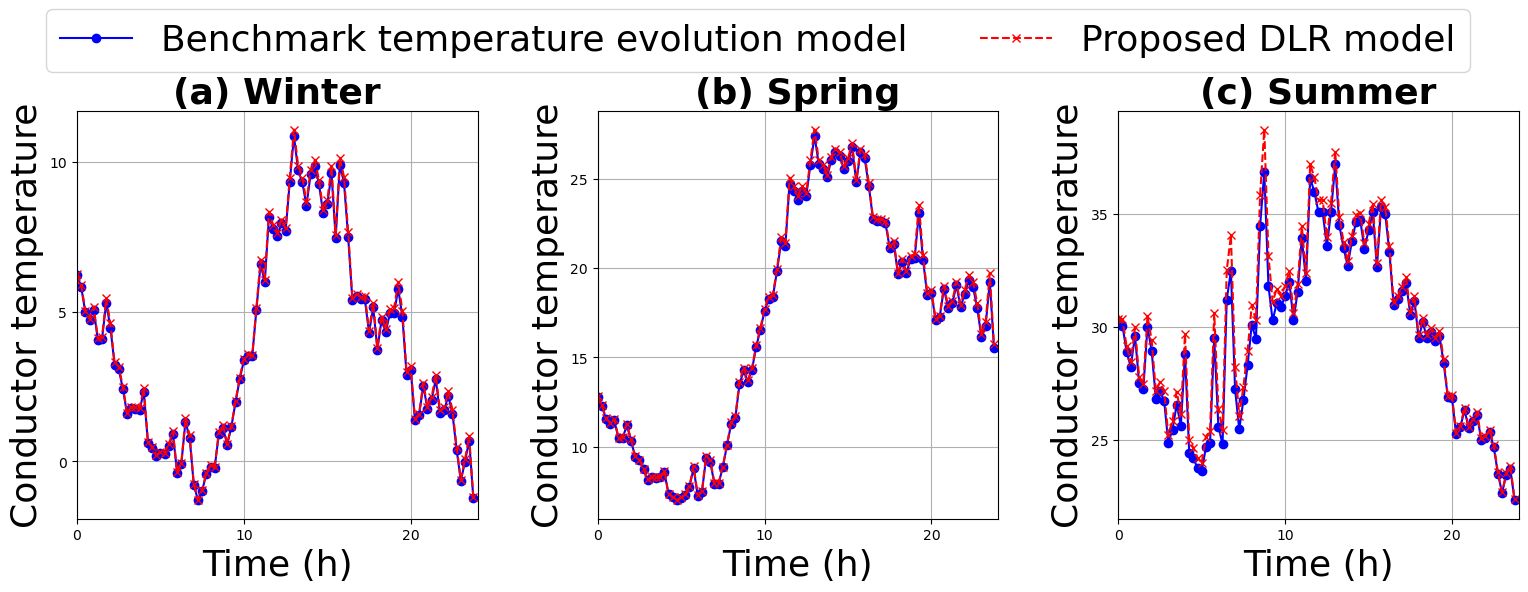}
    \caption{Line temperature evolution under three seasonal scenarios using the benchmark model (red) and proposed model (blue) in \eqref{DLR:transient}}
    \label{fig:test_temp}

\end{figure}

\begin{table}[t]

\caption{The statistic approximation error under three scenarios}
\centering
\begin{tabular}{ccc}
\toprule
Scenario & MAE ($\degree\text{C}$) & Max Error ($\degree\text{C}$) \\
\midrule
Winter & 0.0950 & 0.2484 \\
\midrule
Spring/Fall & 0.1359 & 0.5218 \\
\midrule
Summer & 0.4123 & 1.8684 \\
\bottomrule
\end{tabular}
\label{tab:temp_error}
\end{table}

We analyze a standard ACSR Drake 795 conductor \cite{IEEE:conductor} to compare the proposed approximation in \eqref{DLR:transient} of Theorem 1 and  the benchmark method, which obtains temperature evolution numerically via discrete integration over 1-minute intervals \cite{DLR:bolun},\cite{DLR:TransientAnalysis}. We use weather data from the NREL WIND Toolkit \cite{NREL_WIND_Toolkit} and load data from the NYISO data platform \cite{NYISO_OperationalData}. Using 24-hour data sampled at 15-minute intervals, we test typical winter, fall/spring, and summer conditions with varying wind speed/direction, solar radiation, ambient temperatures, and power flows.  Fig. \ref{fig:test_temp} displays the comparison, in which our method consistently predicts higher temperatures than the benchmark model, validating that it is a tight and conservative bound.  Table \ref{tab:temp_error} also presents the mean absolute error (MAE) and the maximum error.  The summer scenario exhibits the largest MAE (0.4123°C) and the peak error of 1.8684°C at 9 AM, when a relatively high solar radiation coincides with an unusually low wind speed of 1.2 m/s. This is consistent with  (\ref{Approxi:deltars}), where terms $\Delta r = \Delta s = 0$ introduce a linearization error amplified by the inverse of convective heat losses (dominated by wind speed), making the proposed approximation less accurate at lower wind speeds.

    

\subsection{Illustrative example}
\begin{figure}[t]
    \centering
    \includegraphics[width=\columnwidth]{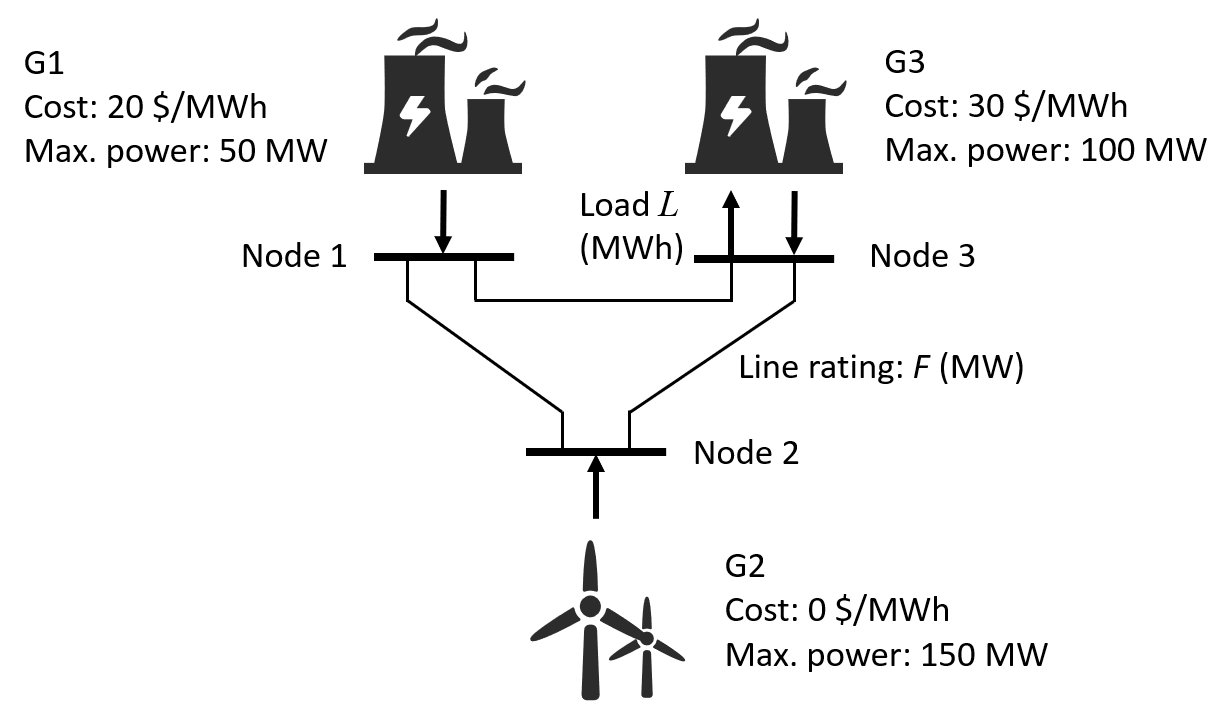}
    \caption{Illustrative three-node system.}
    \label{fig:3_node}
\end{figure}

We first consider a simplified three-node system with two controllable thermal generators and one wind farm as shown in Fig. \ref{fig:3_node}. For simplification, we assume linear production costs for both thermal generators G1 and G3 and zero production costs for wind farm G2. The linear production costs for G1 and G3 are $[20, 30] \$/\text{MWh}$ and the maximum capacity is set to  $[50, 100]$ MW. The wind power forecast is 150 MW. Load $L$ is located at Node 3. The DC approximation is used to model power flow, and we assume that the impedance of the three transmission lines are identical, i.e., $X_{12} = X_{23} = X_{13}$. Furthermore, we assume the capacity of Line 1-2 and Line 1-3 is infinite. We consider a scenario with two time intervals $t_0$ and $t_1$. 
We compare the following three cases with various power flow limit models for Line 2-3: (a) without line flow limits; (b) transient DLRs in \eqref{new_cons}; (c) SLRs of 100 MW.

\begin{figure}[t]
    \centering
    \includegraphics[width=\columnwidth]{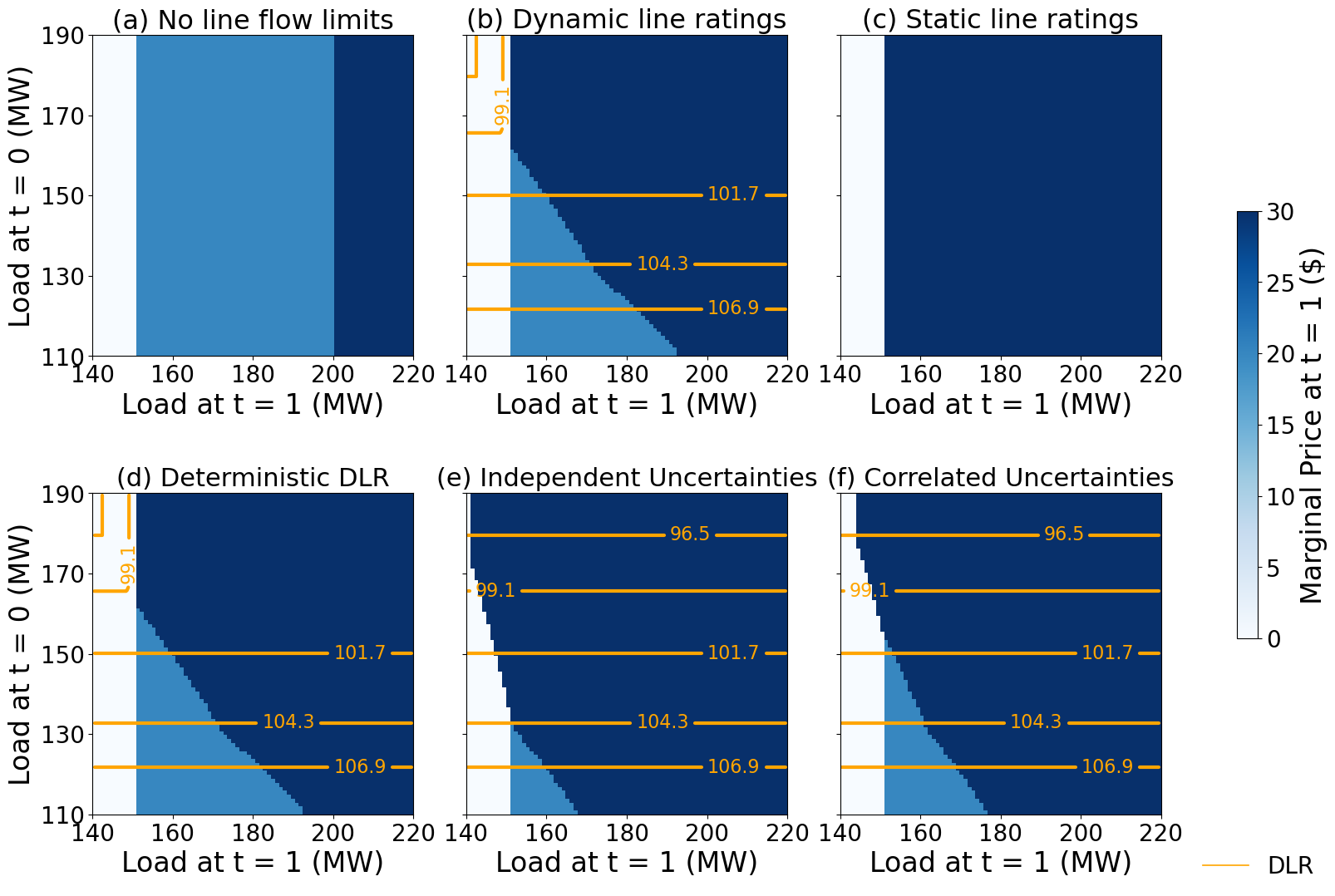}
    \caption{LMP comparison for the illustrative three-node system: cases (a) (b) (c) (d) are conducted under deterministic assumption, where DLR can activate G2 with lower cost as the marginal generator; cases (e) (f) are stochastic model under different uncertainty assumptions which require reserve procurement. The orange contour lines represent the DLR of Line 2-3 at time $t_1$. Correlated uncertainties can moderate this conservativeness compared with independent uncertainties while maintaining system reliability.}
    \label{fig:3_node_deter}
\end{figure}

\begin{figure}[t]
    \centering
    \includegraphics[width=\columnwidth]{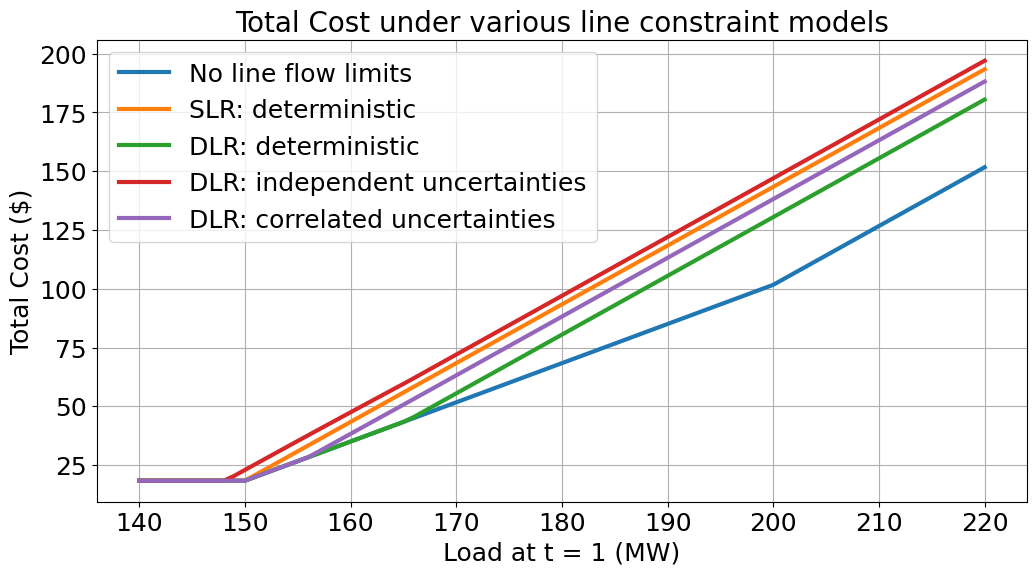}
    \caption{Total cost comparison for the illustrative three-node system}
    \label{fig:3_node_cost}
\end{figure}

Fig. \ref{fig:3_node_deter}(a)-(c) compares LMPs at time $t_1$, given different demand levels for both $t_0$ and $t_1$. The DLR of Line 2-3 at time $t_1$ is determined by the line temperature at the end of period $t_0$ and corresponds to the maximum permissible power flow that would increase the line temperature up to the thermal limit at the end of time interval $t_1$. In the case without line limits, an incremental increase in load $L_1$ at $t_1$ is supplied sequentially by marginal generators G2, G1, and G3, following their economic merit order until each reaches its capacity limit. In  the SLR  case, congestion on Line 2-3 prevents G1 from contributing; consequently, the marginal generator transitions directly from G2 to G3. In the DLR case, a reduction in load $L_0$ at $t_0$ lowers the power flow through Line 2-3, which reduces line temperature at the end of time period $t_0$. The lower initial temperature increases the thermal headroom available in time period $t_1$, thereby raising the effective line rating. As a result, G1 can be activated as a marginal generator to supply additional load at $t_1$, improving the economic efficiency compared to the SLR case (see Fig. \ref{fig:3_node_cost}). In this example, the total cost is bounded between the case without line flow limits and the deterministic SLR case. However, this does not always hold, since DLR can not only expand the feasible operating region (e.g. on windy days), but also can shrink it (e.g. on hot, windless days).

We then conduct a second set of cases, as given in Fig.~\ref{fig:3_node_deter}(d)-(f), to evaluate the impact of different uncertainty models on dispatch outcomes, i.e.,  the deterministic DLR model without uncertainty and reserve requirements,  the  stochastic model with independent uncertainties that require reserve procurement, and the stochastic model with correlated uncertainties in DLR parameters and wind generation. Compared with the deterministic case, both stochastic models lead to greater LMPs for the same load level. This increase is attributed to the explicit allocation of reserves needed to accommodate forecast uncertainty, which imposes additional operational costs on the system. Importantly, the correlation structure in the last case alleviates  this conservativeness. When wind generation and DLR parameters are modeled as positively correlated, the resulting joint uncertainty set is less pessimistic than if these two sources of uncertainty are modeled as  independent. As a result, the correlated case yields lower LMPs.

\begin{figure}[t]
    \centering
    \includegraphics[width=\columnwidth]{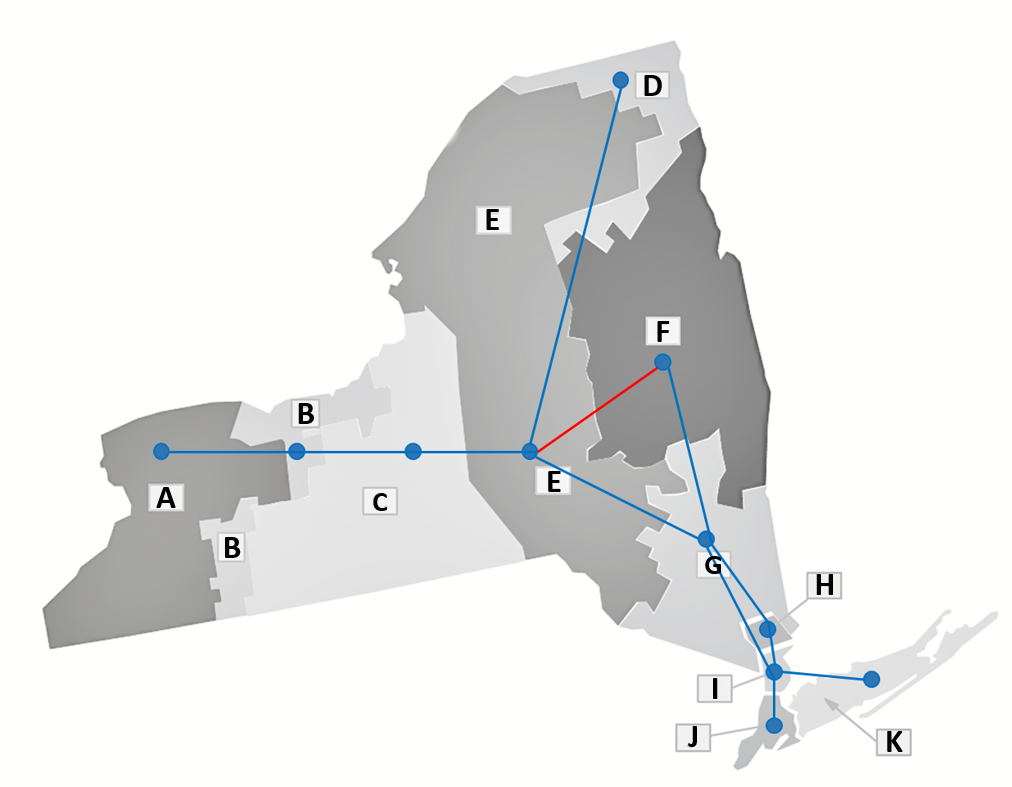}
    \caption{Zonal representations of the NYISO system with DLRs on interface E-F}
    \label{fig:NY_zonal_11}
\end{figure}

\vspace{0.1cm}
\subsection{11-zone NYISO system}\label{4.1}
\subsubsection{Single-day analysis}\label{sub_11_1}
We evaluate the proposed multi-period model in \eqref{MC-OPF} using  the 11-zonal NYISO system in Fig.\ref{fig:NY_zonal_11} \cite{NY11zone}, which includes 12 line interfaces, 361 thermal generators (37.9 GW), and 38 wind farms (6.3 GW primarily in Zones E, G, and K). We model transient DLRs on the historically congested E-F interface for the typical summer day and compare the following three cases: (a) SLRs; (b) DLRs; (c) CC with DLRs and correlated uncertainty.

\begin{figure}[t]
    \centering
    \includegraphics[width=\columnwidth]{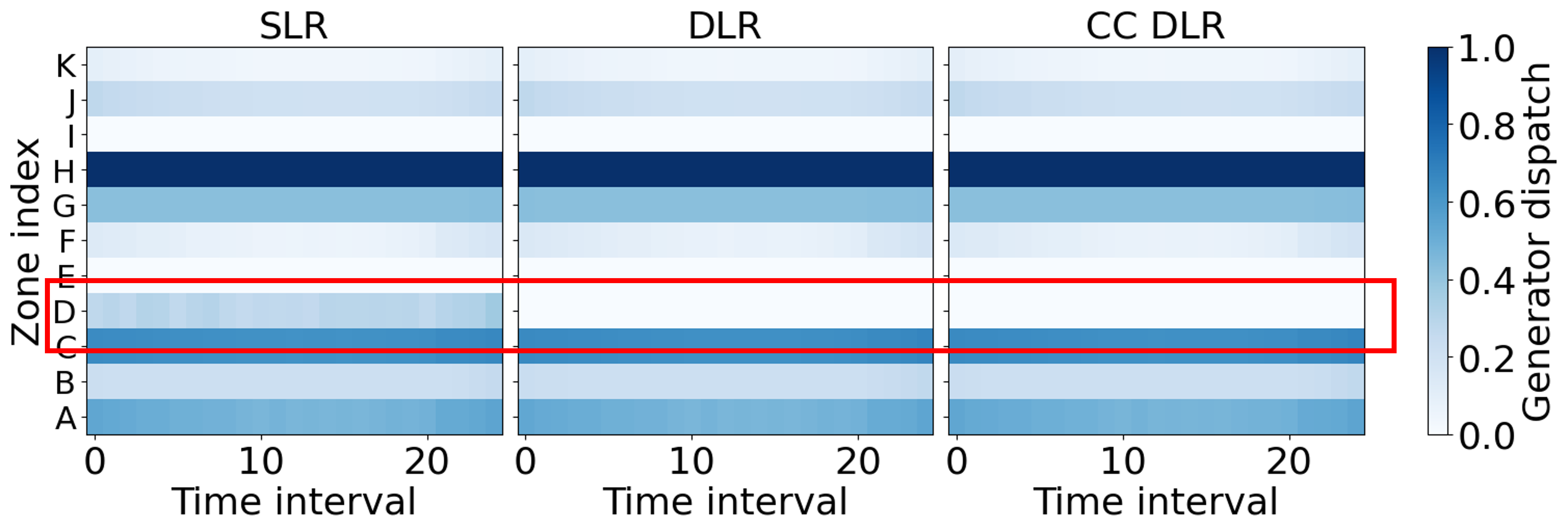}
    \caption{Optimal dispatch results for the 11-zonal NYISO system: both DLR and CC DLR greatly decrease the higher-cost generation in Zone D compared with SLR cases by alleviating the congestion on line E-F (red box)}
    \label{fig:NY_zonal_dispatch}
\end{figure}

\begin{figure}[t]
    \centering
    \includegraphics[width=\columnwidth]{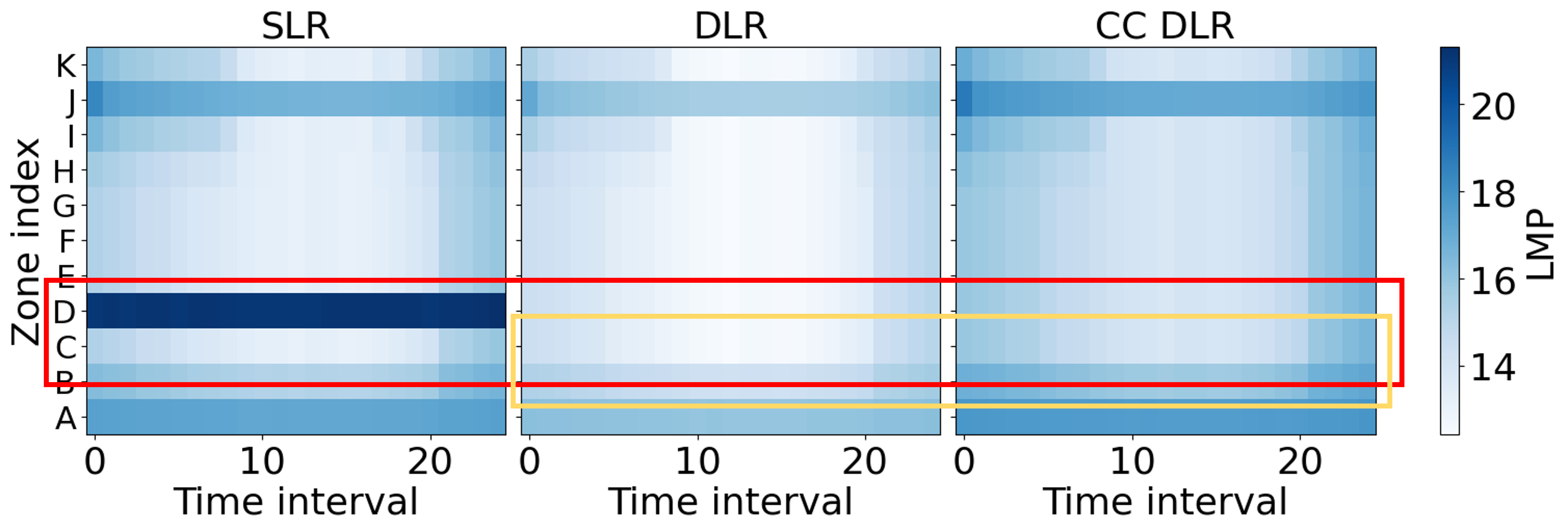}
    \caption{LMP comparison for the 11-zonal NYISO system: DLR reduces the LMP in Zone D at 5:00 from 19.35 \$/MWh to 13.87 \$/MWh (red box). CC DLR increased the LMP in Zone A at 5:00 from 16.07 \$/MWh to 17.63 \$/MWh (yellow box).}
    \label{fig:NY_zonal_LMP}
\end{figure}

\begin{figure}[t]
    \centering
    \includegraphics[width=\columnwidth]{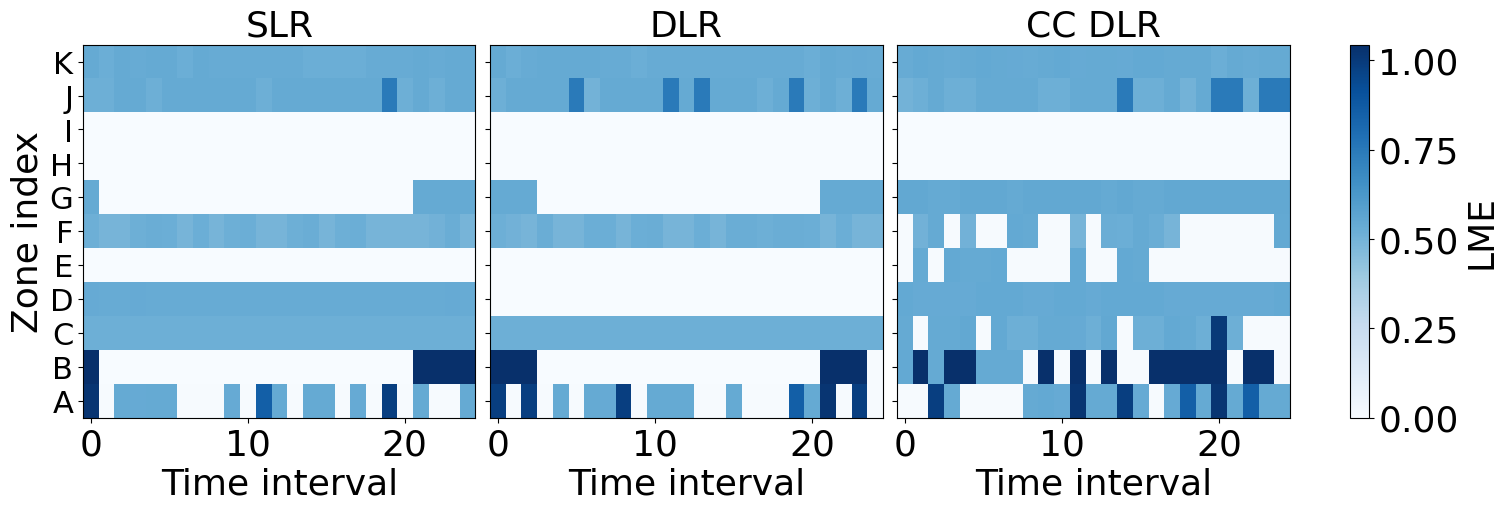}
    \caption{LME comparison for the 11-zonal NYISO system}
    \label{fig:NY_zonal_LME}
\end{figure}

Fig. \ref{fig:NY_zonal_dispatch}--\ref{fig:NY_zonal_LME}   summarize the optimal dispatch, LMP and locational marginal emission (LME)\footnote{We compute LMEs by identifying the marginal generator(s) at each node, i.e., if the output (including reserve) is strictly within  operational limits. The LME is then determined by the emission rate of the marginal generator(s).} results for the three cases. In the SLR case, higher-cost generators in Zone D are marginal due to congestion between zones E--F, preventing the delivery of low-cost electricity from Zone F. DLRs alleviate congestion, reducing more expensive generation in Zone D, leading to the total cost reductions by 0.686\% and  0.242\%  in the DLR and CC-DLR cases. In both DLR cases, DLRs substantially reduce LMP differences between Zones D and E. The CC-DLR case maintains a security margin for weather uncertainty and enables more power transfer than SLR case (by 15.38\%). Considering correlated uncertainty (CC-DLR case) leads to minor dispatch differences with the DLR case, but more pronounced and bidirectional LMP impacts (see Fig. \ref{fig:NY_zonal_LMP}). For instance, the DLR case reduces the LMP in Zone D from 19.35 \$/MWh to 13.87 \$/MWh. However, considering the correlated uncertainty in the CC-DLR case may drive the need for additional transmission reserve margins and, thus,  suppress access to some low-cost generation. For example, this raises the LMP in Zone A at 5:00 from 16.07 \$/MWh to 17.63 \$/MWh. The increasing total cost for the CC-DLR case, as compared with the DLR case, arises from the additional reserve provision, which is 5.14 MW in Zone D at 5:00. The corresponding LMRP is 52.77 \$/MWh, higher than LMP since it reflects the opportunity cost and flexibility premium of keeping capacity available. Both DLR cases also reduce the total carbon emissions and, notably, some LMEs--for instance, in Zone D at 5:00 from 0.537 kg/kWh to 0 kg/kWh. 

\subsubsection{Weighted multi-day analysis}
We use the representative days in Table \ref{tab:temp_error} to extend the analysis in Section \ref{sub_11_1}.  Table \ref{tab:dlr_comparison} summarizes the changes in total system cost and total emissions under DLR and CC-DLR, relative to the baseline case with SLR. 
Under all three scenarios, the introduction of DLR increases the LME in Zone E. This is because the original congestion on interface E-F limits wind power export, resulting in full local consumption. With DLR and CC-DLR, the increased E-F capacity enables abundant wind export, and additional demand in Zone E must be met by carbon-emitting thermal units, thereby raising the LME.

\begin{table}[t]
\centering
\caption{Impacts of DLR and CC DLR on System Cost and Emission Compared to SLR Baseline}
\label{tab:dlr_comparison}
\begin{tabular}{>{\centering\arraybackslash}p{1.2cm}  
                >{\centering\arraybackslash}p{1.2cm}  
                >{\centering\arraybackslash}p{1.2cm}  
                >{\centering\arraybackslash}p{1.2cm}  
                >{\centering\arraybackslash}p{1.2cm}} 
\toprule
\multirow{2}{*}{\textbf{Season}} & \multicolumn{2}{c}{\textbf{Total cost}} & \multicolumn{2}{c}{\textbf{Total emission}} \\
\cmidrule(lr){2-3} \cmidrule(lr){4-5}
& \textbf{DLR} & \textbf{CC DLR} & \textbf{DLR} & \textbf{CC DLR} \\
\midrule
Spring/Fall & -0.67\% & -0.41\% & -0.81\% & -0.38\% \\
Summer & -0.09\% & 0.00\% & -1.58\% & -1.61\% \\
Winter   & -0.78\% & -0.75\% & -2.25\% & -2.23\% \\
\bottomrule
\end{tabular}

\end{table}
\subsection{1814-node NYISO system}


\begin{figure}[t]
    \centering
    \includegraphics[width=\columnwidth]{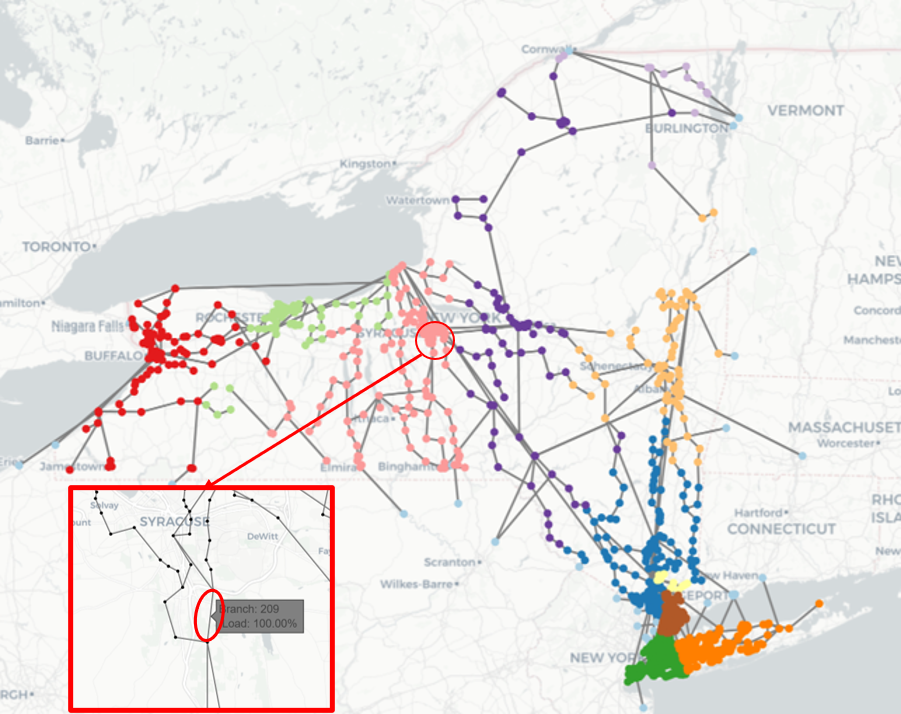}
    \caption{Nodal representations of the NYISO system with DLRs on Line 209}
    \label{fig:NY_nodal_1814}
\end{figure}

\begin{figure}[t]
    \centering
    \includegraphics[width=\columnwidth]{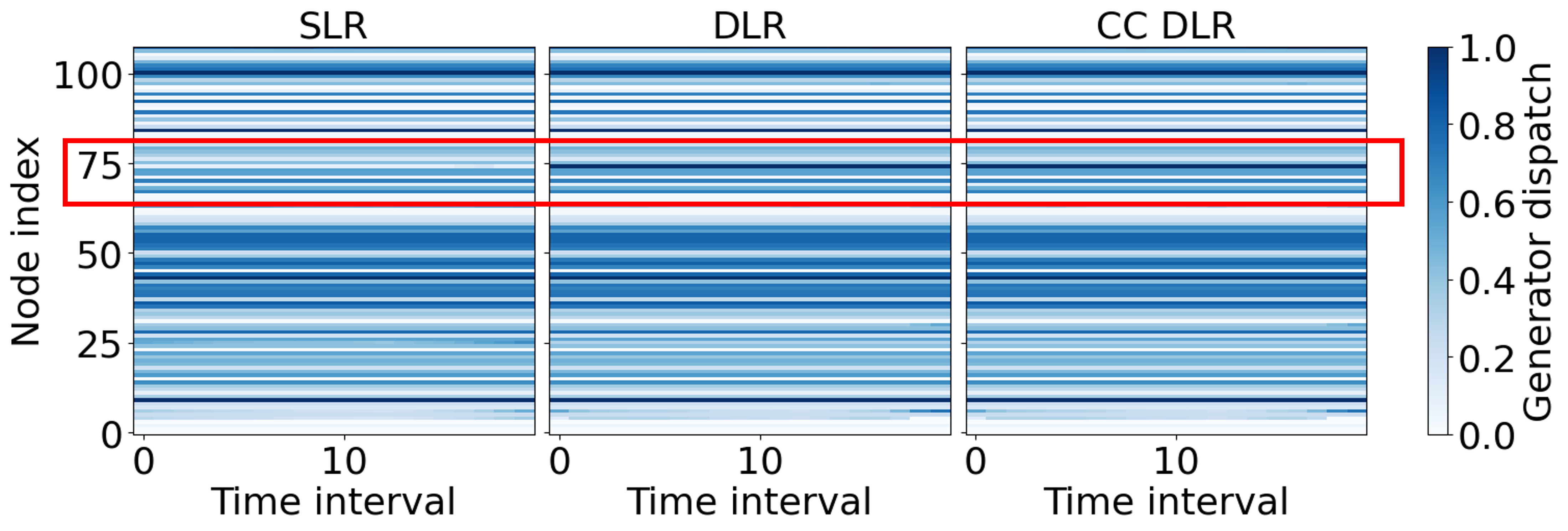}
    \caption{Optimal dispatch results on 106 nodes with generators for the 1814-node NYISO system: the generator at Node 74 increases from 8.9\% to 100\% generation with changes between the SLR and DLR cases (red box).}
    \label{fig:NY_1814_dispatch}
\end{figure}

\begin{figure}[t]
    \centering
    \includegraphics[width=\columnwidth]{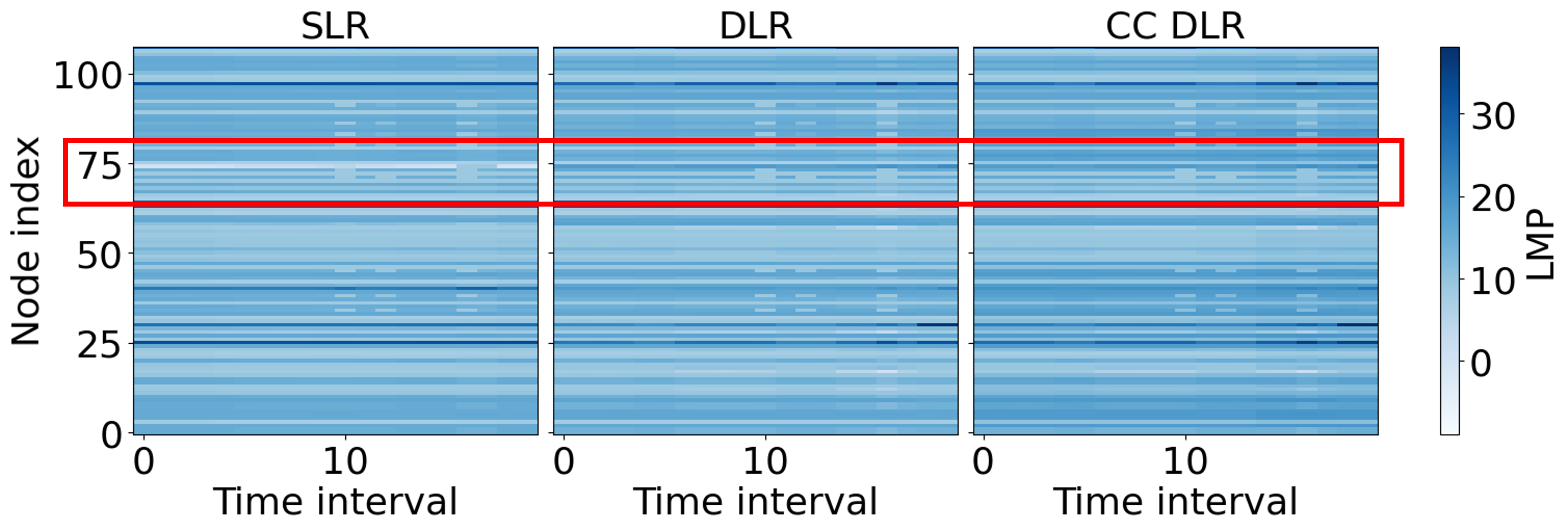}
    \caption{LMPs on 106 nodes with generators for the 1814-node NYISO system: the LMP at Node 74 rises from nearly zero with SLRs to 11.395 \$/MWh with DLRs (red box)}
    \label{fig:NY_1814_LMP}
\end{figure}

\begin{figure}[t]
    \centering
    \includegraphics[width=\columnwidth]{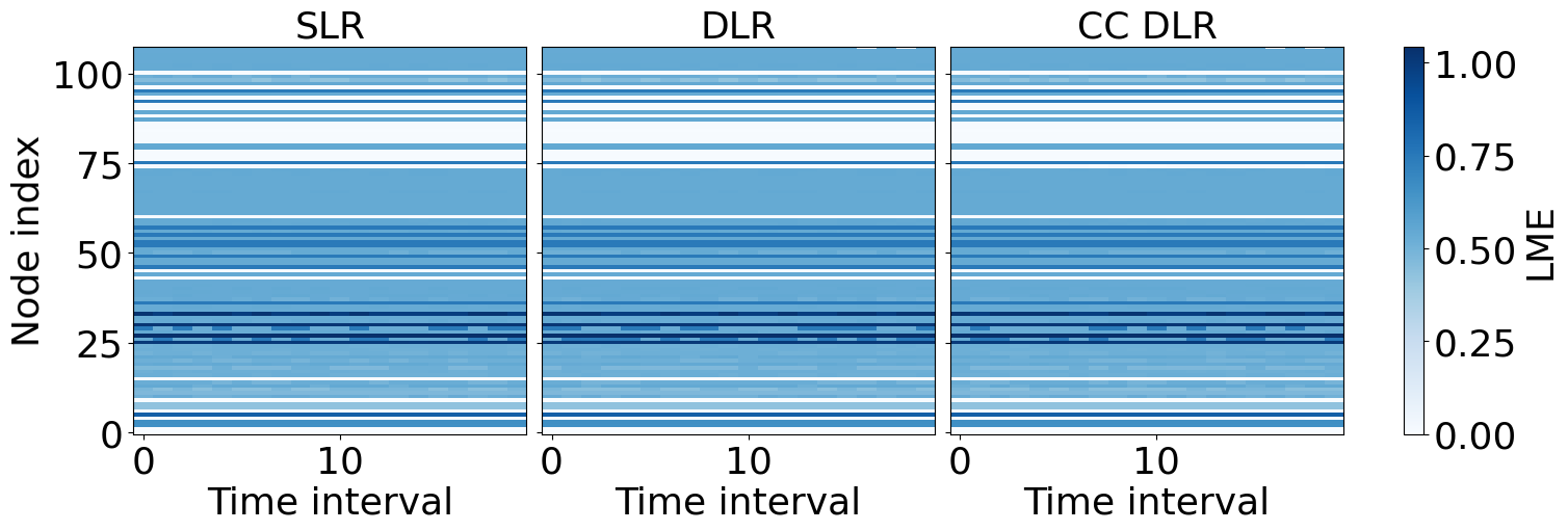}
    \caption{LMEs on 106 nodes with generators for the 1814-node NYISO system}
    \label{fig:NY_1814_LME}
\end{figure}

We use the 1814-node NYISO system \cite{liang2023data} shown in  Fig. \ref{fig:NY_nodal_1814} to evaluate scalability of the proposed model. This system includes 2208 branches, 361 generators (37,852 MW total), and 38 wind farms (6,305 MW total). The structure and parameters of this NYISO system are retrieved from the publicly available data in  \cite{NYISO2022RNA}. All case study data used in this paper are organized in \cite{DLRDataset}. DLRs are modeled for frequently congested Branch 209 using the summer scenario. Fig. \ref{fig:NY_1814_dispatch} shows that DLRs reduce congestion, enabling production from low-cost Generator 71. Fig. \ref{fig:NY_1814_LMP} and \ref{fig:NY_1814_LME} summarize LMPs and LMEs. We list all 106 nodes that host generators out of the total 1814 nodes. LMP differential across the system decreases in the two DLR cases. However, nodal results are again bidirectional, despite the system-wide cost savings and emissions reductions. For example, the LMP at Node 74 rises from nearly zero with SLRs to 11.395 \$/MWh with DLRs, while the LMP at Node 25 decreases during most periods. Similarly, in Fig. \ref{fig:NY_1814_LME}, the LMEs generally reduce under DLRs and CC-DLRs, while the total emissions are reduced by 0.20\% and 0.24\%.


\vspace{0.2cm}
\section{Conclusion}\label{F}
This paper integrates transient DLR models based on the approximate line temperature evolution into electricity market optimization tools, capturing the effects of flexible transmission limits and weather conditions on dispatch and market outcomes. By replacing traditional constraints with transient thermal constraints, the approach enhances transmission utilization while maintaining system security. To solve the multi-period OPF formulation, convexified chance constraints are developed to address correlated uncertainties in DLRs and wind power. The framework introduces thermal reserves to capture uncertainty impacts on transmission capacity and derives LMPs and LMRPs that support competitive market equilibrium.

Our case study shows, compared with SLRs, DLRs can reduce costs by up to 0.686\% (11-zone) and 0.22\% (1814-node) while lowering prices in congested areas by up to 28.3\% and the total emissions by up to 2.23\%. Notably, the choice between zonal and nodal system representations significantly influences cost and emissions outcomes, potentially shaping the deployment and actual benefits of DLRs. Between the zonal and nodal implementations, we do not observe major differences in dispatch decisions, but note that there pronounced impacts in LMPs and LMRPs that may affect revenue opportunities of specific producers.

\bibliographystyle{IEEEtran} 
\bibliography{references}

\appendix
\subsection{Detailed expression for $g(\cdot)$}\label{app:detailed DLR}

1) Solar heat gains are formulated as:
\begin{equation}
    q_s = \alpha_s Q_{s} D,
\end{equation}
where $\alpha_s$ is solar absorptivity, $Q_{s}$ is total solar radiated heat intensity, $D$ is diameter of conductor.

2) Joule heat gains are given by electric power losses as:
\begin{equation}
    q_J = R_c I_c^2 = R_a I_c^2 + \alpha_T  R_{\text{ref}} T_x I_c^2, \label{qj}
\end{equation}
where $R_c$ is conductor resistance at temperature $T_c$. $R_a$ is DC resistance at ambient temperature $T_a$. $\alpha_T$ is temperature coefficient of resistance and $R_{\text{ref}}$ is DC resistance at a reference temperature, $T_{\text{ref}}$. We denote the difference between conductor and ambient temperatures as $T_x = T_c - T_a$.

3) Radiated heat losses are computed as:
\begin{subequations}
    \begin{equation}
    q_r = \pi D h_r T_x
\end{equation}
\begin{equation}
    h_r = \varepsilon \sigma_B (4T_A^3 + 6T_xT_A^2 + 4T_x^2 T_A + T_x^3),
\end{equation}
\end{subequations}
where $T_A = T_a + 273$. $h_r$ is radiative cooling coefficient. $\varepsilon$ is emissivity. $\sigma_B$ is the Stefan-Boltzmann constant.

4) Convective heat losses are given by:
\begin{equation}
    q_c = \pi \lambda_f N_u (T_c - T_a) = \pi D h_c T_x, \label{qc}
\end{equation}
where $\lambda_f$ is thermal conductivity of air. $h_c$ is convective cooling coefficient. $N_u$ is Nusselt number based on wind speed and directions as:
\begin{subequations}
\begin{gather}
    N_u = \max 
    \begin{cases}
         K_{\text{angle}}[1.01+1.35\cdot N_{\text{Re}}^{0.52}] \\
         K_{\text{angle}}\cdot 0.754 \cdot N_{\text{Re}}^{0.6}
    \end{cases}\\
    K_{\text{angle}} \!=\! 1.194 \!-\!\cos{\phi} \!+\! 0.194\cos{2\phi}\!+ \!0.368\sin{2\phi}\\
    N_{\text{Re}} = \frac{\rho_r v D}{v_f},
\end{gather}
\end{subequations}
where $K_{\text{angle}}$ is a wind direction factor, which depends on the angle between the wind direction and the conductor axis $\phi$. $N_{\text{Re}}$ is the Reynolds number. $\rho_r$ is the relative air density compared with that at sea level. $v$ is wind velocity. $v_f$ is the kinematic viscosity.

5) Steady-state dynamic line rating: given the gains and losses above, steady-state DLRs can be computed as:
\begin{subequations}
    \begin{gather}
    q_s + q_J = q_c(T_c^\text{max}) + q_r(T_c^\text{max})\\
    q_s + (R_a + \alpha_TR_{\text{ref}} T_x) (I_c^\text{max})^2 = q_c(T_c^\text{max}) + q_r(T_c^\text{max})\\
    I_c^\text{max} = \sqrt{\frac{q_c(T_c^\text{max})+q_r(T_c^\text{max})-q_s}{R_a + \alpha_T R_{\text{ref}} T_x }}\\
    f_c^\text{max} = g(T_c^\text{max}) = V_c \cdot \sqrt{\frac{q_c(T_c^\text{max})+q_r(T_c^\text{max})-q_s}{R_a + \alpha_T R_{\text{ref}} T_x }},
    \end{gather}
\end{subequations}
where $V_c$ is the voltage magnitude.

All heat balance formulations employed above, including $q_s,q_J,q_c,q_r$, are derived from the IEEE standard \cite{IEEE:conductor} and CIGRE standard \cite{CIGRE207}, which are widely adopted in the literature on DLRs \cite{DLR:wang}, \cite{DLR_kirilenko} and conductor thermal modeling \cite{DLR:TransientAnalysis}. 

\vspace{0.2cm}
\subsection{Proof of Theorem \ref{theorem:DLR-tran}}\label{app:DLR-tran}
Steady-state approximation for DLRs is studied in \cite{DLR_kirilenko} and \cite{DLR:approxi}. Here we extend the basic approximation idea in \cite{DLR:approxi} to transient process.

We first consider the steady-state condition in \eqref{bal} and consider the radiated cooling coefficient $h_r$. Since $T_A \gg T_x$, the last two terms of $h_r$ can be ignored, which is $4 T_x^2 T_A + T_x^3$. Then we define several auxiliary parameters:
\begin{subequations}
    \begin{gather}
    h_{r0} = 4 \sigma_B \varepsilon (T_a + 273)^3\\
    k_1 = 6 \sigma_B \varepsilon(T_a + 273)^2\\
    h'_r = h_{r0} + k_1 T_x \label{h'r}\\
    \Delta r = \pi D \sigma_B \varepsilon(4 T_x^2 T_A + T_x^3).
    \end{gather}
\end{subequations}
Here $h_{r0}$ and $k_1$ are independent of the conductor temperature, and $\Delta r$ is a relatively small positive quantity. Then we compute the radiative cooling coefficient as:
\begin{align}
    q_r & = \pi D h_r T_x \nonumber\\
    &= \pi D \varepsilon \sigma_B(4T_A^3 + 6T_xT_A^2 + 4T_x^2 T_A + T_x^3)T_x \nonumber\\
    &= \pi D (4 \varepsilon \sigma_B T_A^3 \!+\! 6\varepsilon \sigma_B T_A^2T_x)T_x \!+\! \pi D \varepsilon \sigma_B (4T_x^2 T_A \!+\! T_x^3)T_x \nonumber \\
    &= \pi D \left[  4 \varepsilon \sigma_B T_A^3 + 6\varepsilon \sigma_B T_A^2T_x \right] T_x + \Delta rT_x  \label{qr} \\
    &= \pi D(h_{r0}+k_1 T_x)T_x + \Delta r T_x \nonumber\\
    &= (\pi D h'_r + \Delta r)T_x \nonumber.
\end{align}
Furthermore, we set $T_c = T_c^{\text{max}}$ to get the worst-case steady-state heating balance according to \eqref{bal}:
\begin{equation}\label{bal_max}
    q_c(T_c^{\text{max}})+q_r(T_c^{\text{max}})=q_s+q_J(T_c^{\text{max}}, I_c).
\end{equation}
Combining \eqref{bal_max} with \eqref{qj}, \eqref{qc}, \eqref{h'r}, \eqref{qr} leads to:
\begin{subequations}
    \begin{gather}
        q_s \!+\! R_\text{max} I_c^2 \!=\! \pi D h_c T_x \!+\! [\pi D (h_{r0} \!+\! k_1 T_x) \!+\! \Delta r]T_x\\
        R_\text{max} I_c^2 \!+\! q_s \!=\! \pi D[h_c + h_{r0} \!+\! k_1 (T_c^\text{max} \!-\! T_a)] T_x \!+\! \Delta r T_x, \label{Rmax}
    \end{gather}
\end{subequations}
where $R_\text{max}$ is conductor resistance at thermal rating $T_\text{max}$. For simplification, we let $M_{cr} = \pi D[h_c + h_{r0} +k_1 (T_c^\text{max} - T_a)]$. When near thermal rating, Joule heating is considerably more than solar heating, implying that $M_{cr}\gg q_s$. So we let $\Delta s = \frac{q_s}{M_{cr}+\Delta r}  \approx 0$. Then according to \eqref{Rmax}:
\begin{align} 
    T_x &= \frac{R_\text{max} I_c^2 + q_s}{\pi D[h_c + h_{r0} +k_1 (T_c^\text{max} - T_a)] + \Delta r}\nonumber\\
    &= \frac{R_{\text{max}}I_c^2}{M_{cr}+\Delta r} + \frac{q_s}{M_{cr}+\Delta r}\nonumber\\
    &= \frac{R_{\text{max}}}{M_{cr}+\Delta r}I_c^2 +\Delta s \label{k2} \\
    &\approx \frac{R_\text{max}}{M_{cr}} I_c^2 \nonumber\\
    &= k_2 I_c^2, \nonumber
\end{align}
where $k_2 = \frac{R_\text{max}}{M_{cr}}$. Substituting \eqref{qj}, \eqref{qc}, \eqref{qr}, \eqref{k2} into \eqref{bal} leads to the exact formulation:
\begin{subequations}
\begin{equation}
    q_c + q_r = q_s + q_J
\end{equation}
\begin{align}
    [\pi D (h_c +& h_{r0}+k_1T_x) + \Delta r] T_x \nonumber\\
    &= \alpha_s Q_s D + R_a I_c^2 + \alpha_T R_{\text{ref}}I_c^2 T_x
\end{align}
\begin{align}
    [\pi D &(h_c + h_{r0}) + \Delta r] T_x \nonumber\\
    &= \alpha_s Q_s D + R_a I_c^2 + \alpha_T R_{\text{ref}}I_c^2 T_x - \pi D k_1T_x^2 
\end{align}
\begin{align}
    [\pi D (h_c + &h_{r0}) ] T_x = \alpha_s Q_s D + R_a I_c^2 \nonumber\\
    &+ \alpha_T R_{\text{ref}} I_c^2(\frac{R_\text{max}}{M_{cr} + \Delta r}I_c^2 + \Delta s) \label{Delta}\\
    &-\pi D k_1 (\frac{R_\text{max}}{M_{cr} + \Delta r} I_c^2  + \Delta s)^2 -T_x \Delta r. \nonumber
\end{align}
\end{subequations}
We assume the influence of $\Delta r$ and $\Delta s$ in \eqref{Delta} is ignorable. Then we quantify and compare their influence respectively. Define function $F_1(\Delta r, \Delta s)$:
\begin{align}
    F_1(\Delta r, \Delta s) =& \alpha_T R_{\text{ref}} I_c^2(\frac{R_\text{max}}{M_{cr} + \Delta r}I_c^2 + \Delta s) \nonumber \\
    &-\pi D k_1 (\frac{R_\text{max}}{M_{cr} + \Delta r} I_c^2  + \Delta s)^2 - T_x \Delta r \nonumber \\
    =& \tilde{b}_1 I_c^2(\frac{R_\text{max}}{M_{cr} + \Delta r}I_c^2 + \Delta s) \label{F1_detail}\\
    &- \tilde{b}_2 (\frac{R_\text{max}}{M_{cr} + \Delta r}I_c^2 + \Delta s)^2 - T_x \Delta r , \nonumber 
\end{align}
where $\tilde{b}_1 = \alpha_T R_{\text{ref}}$, $\tilde{b}_2 = \pi D k_1$.
Substituting \eqref{F1_detail} into \eqref{Delta}, we have:
\begin{equation}
    \pi D(h_c + h_{r0})T_x = \alpha_s Q_s D + R_a I_c^2 + F_1(\Delta r, \Delta s).
\end{equation}

We then evaluate the case where the ignorable terms $\Delta r$ and $\Delta s$ are set to zero, in order to quantify their actual influence as below:
\begin{equation}
    F_1(\Delta r, \Delta s) < F_1(0,0) = \left[\tilde{b}_1 \frac{R_\text{max}}{M_{cr}} - \tilde{b}_2 (\frac{R_\text{max}}{M_{cr}})^2\right] I_c^4.\label{F1_form}
\end{equation}
To prove \eqref{F1_form}, we apply a first-order Taylor expansion at $(0,0)$:
\begin{equation}
    F_1(\Delta r, \Delta s) \!=\! F_1(0,0) \!+\! \Delta r \frac{\partial F_1(0,0)}{\partial \Delta r} \!+\! \Delta s \frac{\partial F_1(0,0)}{\partial \Delta s}.
\end{equation}

We calculate the derivative of $F_1(\Delta r, \Delta s)$ to $\Delta r$ and $\Delta s$:

\begin{subequations}
\begin{align}
    \frac{\partial F_1(\Delta r, \Delta s)}{\partial \Delta r} =& - \tilde{b}_1 I_c^4  \frac{R_\text{max}}{(M_{cr} + \Delta r)^2} - T_x \\
    &+ 2\tilde{b}_2 I_c^2(\frac{R_\text{max}}{M_{cr} + \Delta r}I_c^2 + \Delta s)\frac{R_\text{max}}{(M_{cr} + \Delta r)^2} \nonumber
\end{align}
\begin{equation}
    \frac{\partial F_1(\Delta r, \Delta s)}{\partial \Delta s} = \tilde{b}_1I_c^2 - 2 \tilde{b}_2 (\frac{R_\text{max}}{M_{cr} + \Delta r}I_c^2 + \Delta s).
\end{equation}
\end{subequations}
Then we can get:
\begin{subequations}\label{Approxi:deltars}
\begin{align}
    \frac{\partial F_1(0,0)}{\partial \Delta r} &= - \tilde{b}_1 I_c^4  \frac{R_\text{max}}{M_{cr}^2} + 2\tilde{b}_2 I_c^4\frac{R_\text{max}^2}{M_{cr}^3} - T_x \\
    \frac{\partial F_1(0,0)}{\partial \Delta s} &= \tilde{b}_1I_c^2 - 2 \tilde{b}_2 I_c^2 \frac{R_\text{max}}{M_{cr}}.
\end{align}
\end{subequations}

So the sufficient condition is:
\begin{equation}\label{proof:static}
    \Delta r \frac{\partial F_1(0,0)}{\partial \Delta r} + \Delta s \frac{\partial F_1(0,0)}{\partial \Delta s}< 0,
\end{equation}
which can be verified by numerical computation. 

Then we obtain:
\begin{align}
    \pi D(h_c + h_{r0})T_x <& \alpha_s Q_s D + R_a I_c^2 \nonumber\\
    &+ \left[\tilde{b}_1 \frac{R_\text{max}}{M_{cr}} - \tilde{b}_2 (\frac{R_\text{max}}{M_{cr}})^2\right] I_c^4,
\end{align}
which decompose the current $I_c$ and the temperature $T_x$ into their respective components.

We then consider to impose transient process. For \eqref{Rmax}, we denote the difference between the transient $T_c$, given $I_c$ and maximum values as $\Delta T = T_c^\text{max} - T_c$, $\Delta I = I_c^\text{max} - I_c$. Then we can reformulate the heating balance \eqref{Rmax} as:
\begin{subequations}
    \begin{align}
        (I_c + \Delta I)^2 R_{\text{max}} + q_s &= (M_{cr} + \Delta r)(T_c^\text{max} - T_a)\\
        &= (M_{cr} + \Delta r)(T_c+\Delta T - T_c + T_x)\nonumber
    \end{align}
    \begin{align}
        T_x &= \frac{R_{\text{max}}(I_c + \Delta I)^2 + q_s}{M_{cr} + \Delta r} - \Delta T\nonumber\\
        &= \frac{R_{\text{max}}}{M_{cr} + \Delta r}(I_c + \Delta I)^2 + \Delta s - \Delta T.\label{Tx_form_dyna}
    \end{align}
\end{subequations}
Substituting \eqref{Tx_form_dyna} into the transient heating balance \eqref{trans}:
\begin{align}
    C\frac{dT_c}{dt} =& q_s + q_J -q_c - q_r \nonumber\\
    =& \alpha_s Q_s D +(R_a+\alpha_T R_{\text{ref}}T_x)I_c^2 \nonumber \\
    &- \pi D h_c T_x - [\pi D (h_{r0}+k_1T_x) + \Delta r]T_x \nonumber\\
    = & \alpha_s Q_s D + R_a I_c^2 - \pi D (h_c + h_{r0}) T_x \nonumber \\
    & + \alpha_T R_{\text{ref}}I_c^2 T_x - \pi D k_1 T_x^2 - \Delta r T_x \\
    = & \alpha_s Q_s D + R_a I_c^2 - \pi D (h_c + h_{r0}) T_x - \Delta r T_x \nonumber \\
    & + \alpha_T R_{\text{ref}}I_c^2 \left[ \frac{R_{\text{max}}}{M_{cr} + \Delta r}(I_c + \Delta I)^2 + \Delta s - \Delta T \right] \nonumber \\
    & - \pi D k_1 \left[ \frac{R_{\text{max}}}{M_{cr} + \Delta r}(I_c + \Delta I)^2 + \Delta s - \Delta T \right]^2 .\nonumber
\end{align}

We disregard the small terms: $\Delta r$, $\Delta s$, $\Delta T$ and $\Delta I$. If \eqref{proof:static} is satisfied, $\Delta r$ and $\Delta s$ can be neglected. Then we consider $\Delta T$, $\Delta I$. Define $F_2(\Delta T,\Delta I)$:
\begin{align}
    F_2(\Delta T,\Delta I) =& \alpha_T R_{\text{ref}}I_c^2 \left[ \frac{R_{\text{max}}}{M_{cr}}(I_c + \Delta I)^2 - \Delta T \right] \nonumber \\
    & - \pi D k_1 \left[ \frac{R_{\text{max}}}{M_{cr}}(I_c + \Delta I)^2 - \Delta T \right]^2\\
    =& \tilde{b}_1 I_c^2 \frac{R_\text{max}}{M_{cr}}(I_c + \Delta I)^2 - \tilde{b}_1 I_c^2 \Delta T \nonumber\\
    &- \tilde{b}_2 \left[ \frac{R_{\text{max}}}{M_{cr}}(I_c + \Delta I)^2 - \Delta T \right]^2. \nonumber
\end{align}

Then we have:
\begin{equation}
    C\frac{dT_c}{dt} \!<\! \alpha_s Q_s D \!+\! R_a I_c^2 \!-\! \pi D (h_c \!+\! h_{r0}) T_x \!+\! F_2(\Delta T, \Delta I).
\end{equation}

We then evaluate the case where the ignorable terms $\Delta T$ and $\Delta I$ are set to zero, in order to quantify their actual influence as below:
\begin{equation}
    F_2(\Delta T, \Delta I) < F_2(0,0) = \left[ \tilde{b}_1 \frac{R_\text{max}}{M_{cr}} - \tilde{b}_2 (\frac{R_\text{max}}{M_{cr}})^2 \right] I_c^4. \label{F2_form}
\end{equation}

To prove \eqref{F2_form}, we conduct a first-order Taylor expansion at $(0,0)$:
\begin{equation}
    F_2(\Delta T, \Delta I) = F_2(0,0) + \Delta T \frac{\partial F_2(0,0)}{\partial \Delta T} + \Delta I \frac{\partial F_2(0,0)}{\partial \Delta I}.
\end{equation}

We calculate the derivative of $F_2(\Delta T, \Delta I)$ to $\Delta T$ and $\Delta I$:
\begin{subequations}
\begin{equation}
    \frac{\partial F_2(\Delta T, \Delta I)}{\partial \Delta T} = \!-\!\tilde{b}_1I_c^2 \!+\! 2 \tilde{b}_2 \left[ \frac{R_{\text{max}}}{M_{cr}}(I_c \!+\! \Delta I)^2 \!-\! \Delta T \right]
\end{equation}
\begin{align}
    \frac{\partial F_2(\Delta T, \Delta I)}{\partial \Delta I} = & 2\tilde{b}_1 I_c^2  \frac{R_\text{max}}{M_{cr}}(I_c+\Delta I) \\
    & - 2\tilde{b}_2 \left[ \frac{R_{\text{max}}}{M_{cr}}(I_c + \Delta I)^2 - \Delta T \right]\frac{R_\text{max}}{M_{cr}}. \nonumber
\end{align}
\end{subequations}
Then we can get:
\begin{subequations}
    \begin{align}
        \frac{\partial F_2(0,0)}{\partial \Delta T} &= -\tilde{b}_1I_c^2 + 2 \tilde{b}_2 I_c^2 \frac{R_{\text{max}}}{M_{cr}}\\
        \frac{\partial F_2(0,0)}{\partial \Delta I} &= 2\tilde{b}_1 I_c^3  \frac{R_\text{max}}{M_{cr}} - 2\tilde{b}_2 I_c^2 \frac{R_{\text{max}}^2}{M_{cr}^2}
    \end{align}
\end{subequations}

So the sufficient condition is:
\begin{equation}\label{proof:trans}
    \Delta T \frac{\partial F_2(\Delta T, \Delta I)}{\partial \Delta T} + \Delta I \frac{\partial F_2(\Delta T, \Delta I)}{\partial \Delta I} < 0.
\end{equation}
Then we have:
\begin{align}
    C\frac{dT_c}{dt} < & \alpha_s Q_s D + R_a I_c^2 - \pi D (h_c + h_{r0}) (T_c-T_a)  \nonumber \\
    & + \left[\tilde{b}_1 \frac{R_{\text{max}}}{M_{cr}} - \tilde{b}_2 (\frac{R_{\text{max}}}{M_{cr}})^2\right]I_c^4.
\end{align}
Furthermore, by discretizing the time intervals, we obtain the conservative temperature evolution:
\begin{equation}\label{appen_appro_DLR}
    T_{t+1} < h(T_t, f_t) = \mu^a_t +\mu^b_t T_t + \mu^c_t f_t^2 + \mu^d_t f_t^4,
\end{equation}
where $\mu^a_t = [\alpha_s Q_s D + \pi D (h_c + h_{r0}) T_a] \frac{\delta t}{mc}$, $\mu^b_t = 1 + \frac{\pi D (h_c + h_{r0})\delta t}{mc}$, $\mu^c_t = \frac{R_a\delta t}{mcV_c^2}$, $\mu^d_t = [\alpha_T R_{\text{ref}} \frac{R_\text{max}}{M_{cr}} - \pi D k_1 (\frac{R_\text{max}}{M_{cr}})^2]\frac{\delta t}{mcV_c^4}$. $\delta t$ is the time interval.

The mathematical approximation error for \eqref{appen_appro_DLR} is:
\begin{align}
        \Delta =& h(T_t, f_t) - T_{t+1} \nonumber \\
        =& T_x \Delta r + \alpha_T R_{\text{ref}}I_c^2 \biggl[ \frac{R_{\text{max}}}{M_{cr}}I_c^2 - \Delta s + \Delta T -\\
        & \frac{R_{\text{max}}}{M_{cr} + \Delta r}(I_c + \Delta I)^2 \biggr]
        - \pi D k_1 \bigg\{  \frac{R_{\text{max}}^2}{M_{cr}^2} I_c^4 - \nonumber\\
        &\biggl[ \frac{R_{\text{max}}}{M_{cr} + \Delta r}(I_c + \Delta I)^2 + \Delta s - \Delta T \biggr]^2   \bigg\}.\nonumber
    \end{align}

\subsection{Proof of Theorem \ref{theorem:market_single}}\label{app:market_single}

For the generator in node-$i$, its objective is to maximize the revenue under energy price $\pi$ and reserve price $\tau$:
\begin{subequations}
\begin{align}
    \max_{\{p_i,R_i^\text{up}, R_i^\text{dn}\}} \quad& \pi p_i+\tau(R_i^\text{up} + R_i^\text{dn}) -c_{i,1}p_i-c_{n,2}p_i^2  \label{GEN:obj}\\
    \quad
    & p_i + R_i^\text{up} \leq p^\text{max}_i \quad:(\nu^+_i) \label{GEN:Pmax}\\
    & p_i - R_i^\text{dn}\geq p^\text{min}_i  \quad:(\nu^-_i), \label{GEN:Pmin}
\end{align}
\end{subequations}
which is to maximize a concave problem. The partial KKT conditions can be calculated as:
\begin{subequations}
    \begin{align}
        (p_i):& -\pi + c_{i,1} + 2c_{i,2}p_i + \nu_i^+ - \nu_i^- =0\label{KKT-CC:gen:p}\\
        (R_i^\text{up}):& -\tau + \nu_i^+ = 0\label{KKT-CC:gen:a}\\
        (R_i^\text{dn}):& -\tau + \nu_i^- = 0.\label{KKT-CC:gen:a-}
    \end{align}
\end{subequations}

The equilibrium for $p_i$ and $\text{LMP}_i$ can be reached when $\pi = \text{LMP}_i$ based on \eqref{LMP:CC}, \eqref{KKT-CC:sys:a}, and \eqref{KKT-CC:gen:a}. The equilibrium for $R_i^\text{up}$, $R_i^\text{dn}$ and $\text{LMRP}_i$ can be reached according to \eqref{LMRP:CC}, \eqref{KKT-CC:sys:a}, \eqref{SOC:LMRP}, \eqref{KKT-CC:gen:a} and \eqref{KKT-CC:gen:a-}.

\vspace{0.2cm}
\subsection{Proof of Theorem \ref{theorem:market_multi}} \label{app:market_multi}

For the generator in node-$i$, its objective is to maximize the revenue under energy price $\pi$ and reserve price $\tau$:
\begin{subequations}
\begin{align}
    \max_{\{p,R^\text{up}, R^\text{dn}\}} &\! \sum_t \pi p_{i,t} \!+\! \tau(R_{i,t}^\text{up} \!+\! R_{i,t}^\text{dn}) \!-\! c_{i,1}p_{i,t} \!-\! c_{i,2}p_{i,t}^2  \label{GEN-multi:obj}\\
    (\nu^+_t):& p_{i,t} + R_{i,t}^\text{up} \leq p^\text{max}_i \label{GEN-multi:Pmax}\\
    (\nu^-_t):& p_{i,t} - R_{i,t}^\text{dn}\geq p^\text{min}_i   \label{GEN-multi:Pmin} \\
    (\eta_t^+):& p_{i,t+1} - p_{i,t} + R^\text{up}_{i,t+1} + R^\text{dn}_{i,t} \leq U_i^\text{up} \label{GEN-multi:UP}\\
    (\eta_t^-):& p_{i,t+1} - p_{i,t} - R^\text{dn}_{i,t+1} - R^\text{up}_{i,t} \geq -U_i^\text{dn},\label{GEN-multi:DN}
\end{align}
\end{subequations}
which is to maximize a concave problem. The partial KKT conditions can be calculated as:
\begin{subequations}
    \begin{align}
        (p_{i,t}):& -\pi + c_{i,1} + 2c_{i,2}p_i + \nu_i^+ - \nu_i^- - \eta_t^+ \nonumber\\
        &\quad+ \eta_t^- + \eta_{t-1}^+ - \eta_{t-1}^-=0\label{KKT-CC:gen-multi:p}\\
        (R_{i,t}^\text{up}):& -\tau + \nu_i^+ + \eta_{t-1}^+ + \eta_t^- = 0\label{KKT-CC:gen-multi:a}\\
        (R_{i,t}^\text{dn}):& -\tau + \nu_i^- + \eta^+_t + \eta^-_{t-1}= 0.\label{KKT-CC:gen-multi:a-}
    \end{align}
\end{subequations}

The equilibrium for $p_{i,t}$ and $\text{LMP}_{i,t}$ can be reached when $\pi = \text{LMP}_{i,t}$ based on \eqref{MC-OPF:LMPa}, \eqref{Multi-KKT:p}, and \eqref{KKT-CC:gen-multi:p}. The equilibrium for $R_{i,t}^\text{up}$, $R_{i,t}^\text{dn}$ and $\text{LMRP}_{i,t}$ can be reached according to \eqref{MC-OPF:LMPb}, \eqref{Multi-KKT:a}, \eqref{MC-OPF:LMRP}, \eqref{KKT-CC:gen-multi:a} and \eqref{KKT-CC:gen-multi:a-}.

\subsection{Relaxation example in a linear two-period system}\label{app:relaxation}
To illustrate the proposed auxiliary reserve-based relaxation in Section \ref{B.3}, we consider a linear CC problem with two periods:
\begin{subequations}\label{test_initial}
    \begin{align}
        \max_{x_1,x_2} \quad & x_1 + 2x_2 + 10 \\
        & \hat{u}_1 = 0\\
        & \hat{u}_2 = 3 x_1 + \frac{1}{3} \hat{u}_1 + \boldsymbol{\varphi}_1 \\
        & \hat{u}_3 = 3 x_2 + \frac{1}{3} \hat{u}_2 + \boldsymbol{\varphi}_2 \\
        & \hat{u}_2 \leq 5\\
        & \hat{u}_3 \leq 5,
    \end{align}
\end{subequations}
where $\boldsymbol{\varphi}_1$ and $\boldsymbol{\varphi}_2$ denote zero-mean Gaussian random variables $\boldsymbol{\varphi}_1 \sim \mathcal{N}(0, \sigma_1)$, $\boldsymbol{\varphi}_2 \sim \mathcal{N}(0, \sigma_2)$. The problem \eqref{test_initial} is equivalent to:
\begin{subequations}\label{test_equal}
    \begin{align}
        \text{Obj}_1 = \nonumber\\
        \max_{x_1,x_2} \quad & x_1 + 2x_2 +10 \\
        & \mathbb{P}_{\varphi_1}[3x_1 + \boldsymbol{\varphi}_1 \leq 5] \geq 1-\epsilon\\
        & \mathbb{P}_{\varphi_1, \varphi_2}[3x_2 + x_1 + \frac{1}{3}\boldsymbol{\varphi}_1+\boldsymbol{\varphi}_2 \leq 5] \geq 1-\epsilon .
    \end{align}
\end{subequations}

After applying the relaxation methods proposed in Section \ref{B.3}, problem \eqref{test_initial} can be reformulated as:
\begin{subequations}\label{test_relax}
    \begin{align}
        \text{Obj}_2 = \max_{x_1,x_2} \quad & x_1 + 2x_2 + 10\\
        & u_2 = 3x_1\\
        & u_3 = 3x_2 + \frac{1}{3}u_2 \\
        & u_2 + R_2 <= 5\\
        & u_3 + R_3 <= 5\\
        & \mathbb{P}_{\varphi_1}[R_2 \geq \boldsymbol{\varphi}_1] \geq 1-\epsilon\\
        & \mathbb{P}_{\varphi_2}[R_3 \geq \frac{1}{3}R_2 +\boldsymbol{\varphi}_2] \geq 1-\epsilon ,
    \end{align}
\end{subequations}
where $R_2$ and $R_3$ are auxiliary reserves for $\hat{u}_2-u_2$ and $\hat{u}_3-u_3$. 

Using the Gaussian quantile, the CCs in \eqref{test_equal} and \eqref{test_relax} can be expressed in deterministic form. For \eqref{test_equal} we obtain:
\begin{subequations}
    \begin{gather}
        3x_1 + z\sigma_1 \leq 5 \\
        3 x_2 + x_1 + z\sqrt{(\frac{\sigma_1}{3})^2 + \sigma_2^2}\leq 5,
    \end{gather}    
\end{subequations}
where $z = \Phi^{-1}(1-\epsilon) \approx 1.645$. While for problem \eqref{test_relax}, by minimizing the auxiliary reserves to their tightest feasible values, we have:
\begin{subequations}
    \begin{gather}
        3x_1 + z\sigma_1 \leq 5\\
        3 x_2 + x_1 + z(\frac{1}{3}\sigma_1 + \sigma_2)\leq 5.
    \end{gather}
\end{subequations}
Both reformulations are linear programs with closed-form optimal values. The optimal objectives are:
\begin{subequations}
    \begin{gather}
        \text{Obj}_1 \!=\! \frac{1}{9}(5\!-\!z\sigma_1)\!+\!\frac{2}{3}\left( 5 \!-\!z\sqrt{(\frac{\sigma_1}{3})^2\!+\!\sigma_2^2}\right)\!+\!10\\
        \text{Obj}_2  = \frac{1}{9}(5-z\sigma_1)+\frac{2}{3}(5-z(\frac{\sigma_1}{3}+\sigma_2))\!+\!10,
    \end{gather}
\end{subequations}
which follows that:
\begin{equation}
    \text{Obj}_1 - \text{Obj}_2 = \frac{2}{3}z\left[ \frac{\sigma_1}{3}+\sigma_2 -\sqrt{(\frac{\sigma_1}{3})^2+\sigma_2^2} \right] \geq 0.
\end{equation}

\begin{figure}[t]
    \centering
    \includegraphics[width=\columnwidth]{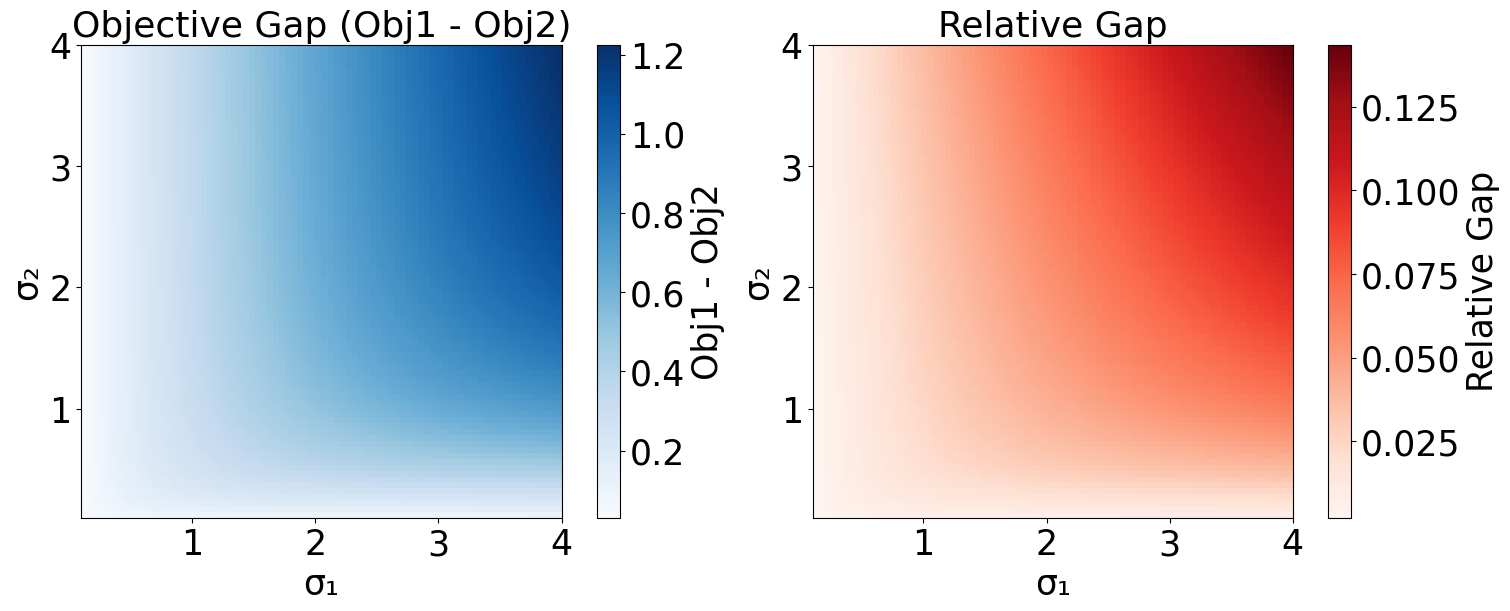}
    \caption{Optimal absolute and relative gap for the proposed auxiliary reserve-based relaxation methods on a linear chance-constrained problem}
    \label{fig:test}
\end{figure}

We evaluate the two formulations on a grid of $\sigma_1,\sigma_2 \in [0.1, 4.0]$. Fig. \ref{fig:test} shows the optimal gap and relative gap of our relaxation. Across the entire range, the optimal objective of \eqref{test_equal} is greater than that of \eqref{test_relax}, consistent with the theoretical bound. As the uncertainty magnitude increases, both formulations produce smaller objectives, and the relaxation gap widens. This reflects the cumulative conservativeness induced by separating uncertainty into auxiliary reserve variables.

\end{document}

%% file: math_and_tools.tex
\usepackage{amsmath,amsfonts,amsthm,amssymb}
\usepackage[bb=boondox]{mathalfa}
\usepackage[algoruled]{algorithm2e}
\usepackage{array}

\usepackage[noabbrev,capitalize]{cleveref}

\crefname{equation}{}{}
\Crefname{equation}{}{}

\usepackage{tikz}

\newenvironment{ldescription}[1]
  {\begin{list}{}%
   {%
    \setlength\labelwidth{2cm}    
    \setlength\leftmargin{\labelwidth}%
    \addtolength\leftmargin{\labelsep}}}
  {\end{list}}

\newcommand{\RNum}[1]{\uppercase\expandafter{\romannumeral #1\relax}}
\newcommand{\rNum}[1]{\expandafter\romannumeral #1}

\makeatletter
\newcommand{\pushright}[1]{\ifmeasuring@#1\else\omit\hfill$\displaystyle#1$\fi\ignorespaces}
\newcommand{\pushleft}[1]{\ifmeasuring@#1\else\omit$\displaystyle#1$\hfill\fi\ignorespaces}
\makeatother

\newcolumntype{C}[1]{>{\centering\arraybackslash}p{#1}} 

%% file: references.bib
@article{DLR:approxi,
  title={Optimal power flow considering line-conductor temperature limits under high penetration of renewable energy},
  author={Ngoko, Bonface O and others},
  journal={Int. J. Electr. Power Energy Syst.},
  volume={101},
  pages={255--267},
  year={2018},
  publisher={Elsevier}
}

@article{DLR:PierrePinson,
  title={Chance-constrained optimal power flow with non-parametric probability distributions of dynamic line ratings},
  author={Viafora, Nicola and others},
  journal={Int. J. Electr. Power Energy Syst.},
  volume={114},
  pages={105389},
  year={2020},
  publisher={Elsevier}
}

@article{DLR:wang,
  title={Risk-based distributionally robust optimal power flow with dynamic line rating},
  author={Wang, Cheng and others},
  journal={IEEE Trans. Power Syst.},
  volume={33},
  number={6},
  pages={6074--6086},
  year={2018},
  publisher={IEEE}
}

@inproceedings{DLR:bolun,
  title={Impacts of dynamic line rating on power dispatch performance and grid integration of renewable energy sources},
  author={Xu, Bolun and others},
  booktitle={IEEE PES ISGT Europe 2013},
  pages={1--5},
  year={2013},
  organization={IEEE}
}

@article{DLR:TransientAnalysis,
  title={Contingency analysis considering the transient thermal behavior of overhead transmission lines},
  author={Wang, Mengxia and others},
  journal={IEEE Trans. Power Syst.},
  volume={33},
  number={5},
  pages={4982--4993},
  year={2018},
  publisher={IEEE}
}

@article{CC:paper1,
  title={Chance-constrained optimal power flow: Risk-aware network control under uncertainty},
  author={Bienstock, Daniel and others},
  journal={Siam Review},
  volume={56},
  number={3},
  pages={461--495},
  year={2014},
  publisher={SIAM}
}

@ARTICLE{CC:paper3,
  author={Halilbašić, Lejla and others},
  journal={IEEE Trans. Power Syst.}, 
  title={Convex Relaxations and Approximations of Chance-Constrained AC-OPF Problems}, 
  year={2019},
  volume={34},
  number={2},
  pages={1459-1470},
  doi={10.1109/TPWRS.2018.2874072}}

@article{Group:tomas,
  title={Electricity Market-Clearing With Extreme Events},
  author={Tapia, Tomas and others},
  journal={IEEE Trans. Energy Markets Policy Regul.},
  year={2024},
  publisher={IEEE}
}

@article{Group:yury,
  title={A chance-constrained stochastic electricity market},
  author={Dvorkin, Yury},
  journal={IEEE Trans. Power Syst.},
  volume={35},
  number={4},
  pages={2993--3003},
  year={2019},
  publisher={IEEE}
}

@article{Report:CongestionCost2,
  author      = {Noah Shreve and others},
  title       = {2023 Transmission Congestion Report},
  institution = {Grid Strategies LLC},
  year        = {2024},
  url         = {http://bit.ly/3ZCrMao}
}

@misc{Report:CAISOcurtail,
  author       = {{California ISO}},
  title        = {Managing the Evolving Grid},
  howpublished = {https://tinyurl.com/5xm75wvb}
}

@article{EPRIreport,
  title={Dynamic line rating systems for transmission lines: Topical report},
  author={Wang, W and Pinter, S},
  journal={US Department of Energy},
  year={2014}
}

@techreport{DLR:DOEreport,
  author       = {Oncor Electric Delivery Company},
  title        = {Oncor’s Dynamic Line Rating (DLR) Project},
  institution  = {U.S. Department of Energy},
  year         = {2014},
  month        = {May},
  url          = {https://tinyurl.com/bdzh4u3a}
}

@techreport{DLR:NYPA,
  author       = {Chunchuan (Charlie) Xu},
  title        = {NYPA Experience with Dynamic Line Rating Technology},
  institution  = {New York Power Authority},
  year         = {2019},
  month        = {September},
  url          = {https://tinyurl.com/y5wdp5yr}
}

@electronic{DLR:EU,
  author       = {{ENTSO-E}},
  title        = {Dynamic Line Rating (DLR)},
  url = {https://tinyurl.com/msnu7hrb}
}

@article{DLR:hard1_hardware:,
  title={Review of dlr systems for wind power integration},
  author={Fernandez, E and others},
  journal={Renew. Sustain. Energy Rev.},
  volume={53},
  pages={80--92},
  year={2016},
  publisher={Elsevier}
}

@inproceedings{DLR:hard3_market,
  title={FERC Order 881-A Step Towards Dynamic Line Ratings for Improved Market Efficiency},
  author={Cheung, Kwok W},
  booktitle={2022 IEEE PES 14th Asia-Pac. Power Energy Eng. Conf. (APPEEC)},
  pages={1--6},
  year={2022},
  organization={IEEE}
}

@article{DLR:hard2_reliability,
  title={DLR: A PRAGMATIC APPROACH TO IMPLEMENTATION},
  author={Wilke, Amy},
  year={2024}
}

@techreport{IEEE:conductor,
  title        = {{IEEE Standard for Calculating the Current-Temperature Relationship of Bare Overhead Conductors}},
  author       = {{IEEE Power and Energy Society}},
  number       = {738-2012},
  type         = {IEEE Standard},
  year         = {2013},
  month        = {Dec},
  doi          = {10.1109/IEEESTD.2013.6692858}
}

@article{NY11zone,
  title={Risk-informed participation in T\&D markets},
  author={Khan, Hafiz Anwar Ullah and others},
  journal={Electr. Power Syst. Res.},
  volume={202},
  pages={107624},
  year={2022},
  publisher={Elsevier}
}

@article{pricing3,
  title={Principles of modern electricity pricing},
  author={Munasinghe, Mohan},
  journal={Proceedings of the IEEE},
  volume={69},
  number={3},
  pages={332--348},
  year={1981},
  publisher={IEEE}
}

@article{scenario1,
  title={Reserve requirements for wind power integration: A scenario-based stochastic programming framework},
  author={Papavasiliou, Anthony and others},
  journal={IEEE Trans. Power Syst.},
  volume={26},
  number={4},
  pages={2197--2206},
  year={2011},
  publisher={IEEE}
}

@article{robust1,
  title={Adaptive robust optimization for the security constrained unit commitment problem},
  author={Bertsimas, Dimitris and others},
  journal={IEEE Trans. Power Syst.},
  volume={28},
  number={1},
  pages={52--63},
  year={2012},
  publisher={IEEE}
}

@article{wind_gen,
  title={A parametric model for wind turbine power curves incorporating environmental conditions},
  author={Saint-Drenan and others},
  journal={Renewable Energy},
  volume={157},
  pages={754--768},
  year={2020},
  publisher={Elsevier}
}

@inproceedings{liang2023data,
  title={Data-driven inverse optimization for marginal offer price recovery in electricity markets},
  author={Liang, Zhirui and others},
  booktitle={Proceedings of the 14th ACM International Conference on Future Energy Systems},
  pages={497--509},
  year={2023}
}

@techreport{DOE2019DLR,
  title        = {{Dynamic Line Rating Report to Congress}},
  institution  = {{U.S. Department of Energy}},
  series       = {Report to Congress},
  month        = jun,
  year         = 2019,
  address      = {Washington, DC},
  url          = {https://tinyurl.com/mmak8v7a},
  note         = {DOE response to H. Rep.\ 115-697 and H. Rep.\ 115-929},
}

@techreport{SLR_cal2,
  author       = {{Idaho National Laboratory}},
  title        = {{Variable Transmission Line Ratings}},
  institution  = {Idaho National Laboratory},
  year         = 2024,
  month        = mar,
  address      = {Idaho Falls, ID},
  url          = {https://tinyurl.com/2dy2ucj3},
}

@article{DLR_kirilenko,
  title={Risk-averse stochastic dynamic line rating models},
  author={Kirilenko, Aleksei and Esmaili, Masoud and Chung, CY},
  journal={IEEE Transactions on Power Systems},
  volume={36},
  number={4},
  pages={3070--3079},
  year={2020},
  publisher={IEEE}
}

@techreport{CIGRE207,
  author       = {{CIGR{\'E} Working Group 22.12}},
  title        = {Thermal Behaviour of Overhead Conductors},
  institution  = {CIGR{\'E} Technical Brochure No.~207},
  year         = {2002},
  address      = {Paris, France}
}

@article{DLR:forecast,
  title={Forecasting for dynamic line rating},
  author={Michiorri, Andrea and Nguyen, Huu-Minh and Alessandrini, Stefano and Bremnes, John Bj{\o}rnar and Dierer, Silke and Ferrero, Enrico and Nygaard, Bj{\o}rn-Egil and Pinson, Pierre and Thomaidis, Nikolaos and Uski, Sanna},
  journal={Renewable and sustainable energy reviews},
  volume={52},
  pages={1713--1730},
  year={2015},
  publisher={Elsevier}
}

@misc{CAISO15min_DA,
  title = {CAISO Market Enhancements: 15-minute Scheduling Option},
  howpublished = {CAISO Market Rule Change Proposal (FERC Order 764 Compliance)},
  year = {2017}
}

@misc{FERCOrder825_RT,
  title = {FERC Order No. 825: Aligning Dispatch and Settlement Intervals},
  howpublished = {FERC Docket No. ER17-772-000},
  note = {Requires 5-minute settlement and dispatch for real-time markets},
  year = {2016}
}

@misc{FERCOrder881,
  author       = {{Federal Energy Regulatory Commission (FERC)}},
  title        = {Order No. 881: Managing Transmission Line Ratings},
  howpublished = {\url{https://www.ferc.gov/media/order-no-881-managing-transmission-line-ratings}},
  year         = {2021},
  note         = {Issued December 16, 2021}
}

@techreport{INL_ThermalInertia,
  author       = {{Idaho National Laboratory}},
  title        = {Dynamic Thermal Circuit Ratings and their Application in Grid Operations},
  institution  = {Idaho National Laboratory},
  year         = {2015},
  url          = {https://www.ferc.gov/sites/default/files/2020-09/Gentle-INL.pdf},
  note         = {FERC Report, accessed 2025}
}

@article{DLR:transient2,
  title={Conductor temperature estimation and prediction at thermal transient state in dynamic line rating application},
  author={Alvarez, David L and da Silva, Filipe Faria and Mombello, Enrique E and Bak, Claus Leth and Rosero, Javier A},
  journal={IEEE Transactions on Power Delivery},
  volume={33},
  number={5},
  pages={2236--2245},
  year={2018},
  publisher={IEEE}
}

@techreport{Abboud2021ConcurrentCooling,
  author       = {Abboud, Alexander W. and Phillips, Tyler and Lehmer, Jacob P. and Starks, Brandon and Gentle, Jake P.},
  title        = {{Dynamic Line Rating Study of Concurrent Cooling for a Proposed Wind Farm}},
  institution  = {Idaho National Laboratory},
  year         = {2021},
  month        = mar,
  number       = {INL-EXT-21-62113, Revision 0},
  url          = {https://inldigitallibrary.inl.gov/sites/sti/sti/Sort\_37483.pdf},
  note         = {Idaho National Laboratory report prepared for U.S. Department of Energy Wind Energy Technologies Office}
}

@techreport{GentleDLRForecasting,
  author       = {Gentle, Jake P. and others},
  title        = {{Forecasting Dynamic Line Ratings with Spatial Variation}},
  institution  = {Idaho National Laboratory},
  year         = {2023},
  url          = {https://www.ferc.gov/sites/default/files/2020-09/Gentle-INL.pdf},
  note         = {INL presentation at FERC DLR Workshop, Sept. 10, 2019}
}

@misc{NYISO2022RNA,
  author       = {{New York Independent System Operator (NYISO)}},
  title        = {2022 Reliability Needs Assessment},
  institution  = {NYISO},
  year         = {2022},
  note         = {Comprehensive evaluation of bulk power system reliability in New York},
  url          = {https://www.nyiso.com/documents/20142/2248793/2022-RNA-Report.pdf}
}

@misc{NREL_WIND_Toolkit,
  author       = {{National Renewable Energy Laboratory (NREL)}},
  title        = {NREL Wind Integration National Dataset (WIND) Toolkit},
  year         = {2015},
  howpublished = {\url{https://www.nrel.gov/grid/wind-toolkit.html}},
  note         = {Meteorological, power, and forecast datasets at high spatial (2 km) and temporal (5 min–hourly) resolution for the continental U.S.},
}

@misc{NYISO_OperationalData,
  author       = {{New York Independent System Operator (NYISO)}},
  title        = {Energy Market \& Operational Data},
  year         = {2025},
  howpublished = {\url{https://www.nyiso.com/energy-market-operational-data}},
  note         = {Provides real-time and day-ahead market data including locational prices, load, fuel mix, and flow metrics}
}

@article{bienstock2014chance,
  title={Chance-constrained optimal power flow: Risk-aware network control under uncertainty},
  author={Bienstock, Daniel and Chertkov, Michael and Harnett, Sean},
  journal={Siam Review},
  volume={56},
  number={3},
  pages={461--495},
  year={2014},
  publisher={SIAM}
}

@article{attack,
  title={Transmission line rating attack in two-settlement electricity markets},
  author={Ye, Hongxing and Ge, Yinyin and Liu, Xuan and Li, Zuyi},
  journal={IEEE Transactions on Smart Grid},
  volume={7},
  number={3},
  pages={1346--1355},
  year={2015},
  publisher={IEEE}
}

@article{Graf25,
    author = {Christoph Graf},
    title  = {Simplified Short-Term Electricity Market Designs: {E}vidence from {E}urope},
    journal   = {The Electricity Journal},
    year   = {2025},
    pages={107486},
    doi={https://doi.org/10.1016/j.tej.2025.107486}
}

@misc{DLRDataset,
  author       = {Zhiyi Zhou},
  title        = {DLR Dataset},
  year         = {2025},
  howpublished = {\url{https://github.com/Zhiyi-miyo/DLR\_Dataset}},
  note         = {Accessed: Sep. 19, 2025}
}
